\documentclass[aps,prb,twocolumn,amsfonts,amsmath,amssymb,superscriptaddress,showpacs,]{revtex4-1}

\usepackage{graphicx}
\usepackage[utf8]{inputenc}
\usepackage{subfigure}
\usepackage{listings}
\usepackage{bm}
\usepackage{setspace}
\usepackage{hyperref}
\usepackage{natbib}
\usepackage[normalem]{ulem}
\usepackage{epstopdf}
\usepackage{subfigure}
\usepackage{dsfont}
\usepackage[bbgreekl]{mathbbol}
\usepackage{amsfonts}

\def\e{\mathrm{e}}
\def\d{\mathrm{d}}
\def\i{\mathrm{i}}

\makeatother
\begin{document}

\title{ Numerical calculation of spectral functions of the Bose-Hubbard
model using B-DMFT }

\author{Jaromir Panas}
\affiliation{ Institute of Theoretical Physics, Faculty of Physics, University of Warsaw, Pasteura 5, 02-093 Warszawa, Poland }

\author{Anna Kauch}
\affiliation{ Institute of Physics, Academy of Sciences of the Czech Republic, Na Slovance 2, 18221 Praha, Czech Republic }

\author{Jan Kune\v{s}}
\affiliation{ Institute of Physics, Academy of Sciences of the Czech Republic, Na Slovance 2, 18221 Praha, Czech Republic }

\author{Dieter Vollhardt}
\affiliation{Theoretical Physics III, Center for Electronic Correlations and Magnetism, Institute of Physics, University of Augsburg, D-86135 Augsburg, Germany}

\author{Krzysztof Byczuk}
\affiliation{ Institute of Theoretical Physics, Faculty of Physics, University of Warsaw, Pasteura 5, 02-093 Warszawa, Poland }

\date{\today}

\begin{abstract}
We calculate the momentum dependent spectral function of the Bose-Hubbard model on a simple cubic lattice in three dimensions within the bosonic dynamical mean-field theory (B-DMFT). The continuous-time quantum Monte Carlo method is used to solve the self-consistent B-DMFT equations together with the maximum entropy method for the analytic continuation to real frequencies. Results for weak, intermediate, and strong interactions are presented. In the limit of weak and strong interactions very good agreement with results obtained by perturbation theory is found. By contrast, at intermediate interactions the results differ significantly, indicating that in this regime perturbative methods fail do describe the dynamics of interacting bosons.
\end{abstract}


\maketitle


\section{Introduction}

During the last few years impressive progress was made in the experimental investigation of ultracold atomic gases in optical lattices.\cite{bloch_short,bloch_long} It is now possible to measure not only density profiles and static correlation functions but even dynamical quantities such as spectral functions of the trapped particles, using Bragg spectroscopy.\cite{bragg_rmp_2005} Thereby momentum resolved bosonic spectral functions were obtained in the case of condensed bosons\cite{sengstock_nphys,bragg_prl_2008,bragg_3d_njp_2010} and across the phase transition to a Mott insulator.\cite{bragg_MI_prl_2009, bragg_fabbri_prl_2012} By contrast, only rather few calculations of bosonic spectral functions have been performed so far for the Bose-Hubbard model. Early studies were based on the strong-coupling approximation to the Bose-Hubbard model.\cite{sengupta} Later the weak-coupling limit at zero temperature was thoroughly analyzed with the functional renormalization group.\cite{dupuis_prl_2009,dupuis_pra_2009,
kopietz_prl_2009,kopietz_pra_2010} Other approaches include the variational cluster approximation,\cite{knap_sf_2011} which was used to investigate systems in one\cite{knap_1d_2010} and two\cite{knap_2d_2010,knap_sf_2011}  dimensions, the quantum rotor approximation,\cite{zaleski} which was recently applied to three-dimensional systems,\cite{zaleski3d} and the linked-cluster expansion (LCE).\cite{kauch_lce} In particular the LCE, which was employed to approximately solve the equations of the bosonic dynamical mean-field theory (B-DMFT),\cite{byczuk_bdmft} allows one to study the Bose-Hubbard model for strong interactions and near the phase boundary between the superfluid and the Mott-insulating phase. The B-DMFT treats local correlations in time exactly and includes spatial correlations on the mean-field level. It is applicable for all values of the Hubbard interaction, density and temperature. The B-DMFT was derived subsequent to the DMFT for  lattice fermions\cite{metz_voll,kotliar_rmp} and was applied to 
various bosonic problems\cite{anders_dmft,hofstetter_2species_2009} as well as to mixtures\cite{byczuk_bf,anders_bf} of bosons and fermions. It has been extended and applied also to inhomogeneous situations\cite{hofstetter_2species_trap_2011} and to bosonic systems in non-equilibrium\cite{eckstein_noneq_2014}.

In this paper we solve the Bose-Hubbard model within the B-DMFT framework on a simple cubic lattice in three dimensions using a continuous-time quantum Monte Carlo (CT-QMC) solver.\cite{gull_ctqmc} This approach is known to give excellent results for the phase diagram and static properties\cite{anders_dmft_short,anders_dmft} of interacting bosons. We compute the momentum resolved and the momentum integrated spectral functions and the dispersion relation of interacting bosons in, both, the superfluid and the Mott insulating phase. The following questions will be addressed and answered:
(i) How well can strong-coupling approaches, which are known to capture the phase diagram of correlated bosons very well, describe dynamical properties such as spectral functions?
(ii) How does the presence of the superfluid influence the spectral properties of normal bosons?
(iii) How is the dispersion relation modified by the interaction?

The paper is organized as follows: In Section \ref{sect2} we introduce the Bose-Hubbard model  and recapitulate the main steps of the B-DMFT framework. We also introduce an improved method for calculating the self-energy which makes use of two-particle Green functions and discuss the method for the numerical analytic continuation. In Section \ref{sect3} we present the results for the momentum resolved spectral function in the limit of weak, intermediate, and strong interactions, respectively. Finally, Section \ref{sect4} concludes the paper with a summary.


\section{Model and investigation method}\label{sect2}



We consider spinless bosons on a lattice described by the Bose-Hubbard Hamiltonian
\begin{equation}\label{bh_ham}
\hat{H}=-\sum_{ij} t_{ij} \hat{b}^{\dag}_{i} \hat{b}_j - \mu \sum_{i} \hat{b}^{\dag}_{i} \hat{b}_i  + \frac{U}{2}\sum_{i} \hat{b}^{\dag}_i \hat{b}^{\dag}_i \hat{b}_i \hat{b}_i,
\end{equation}
where $ \hat{b}^{\dag}_i $ ($ \hat{b}_i $) is a bosonic creation (annihilation) operator on a lattice site $ i $, $\mu$ is the chemical potential, $U$ is the local interaction strength, and $ t_{ij} $ is the hopping amplitude. We assume nearest neighbour (NN) hopping, i.e., $t_{ij}=t>0$ if sites $i$ and $j$ are NN and $0$ otherwise. Our calculations were performed for a simple cubic lattice with coordination number $z=6$ and $t=0.5$ in arbitrary units.


\subsection{The B-DMFT action}

In the B-DMFT the lattice problem is replaced by a single-site (``impurity'') problem with self-consistency conditions. A detailed derivation can be found in Ref. \onlinecite{byczuk_bdmft}. The impurity action reads
\begin{equation}\label{loc_action}
\begin{aligned}
S_{loc}= & \int_0^{\beta} \d \tau b^{\ast}(\tau)\left(\partial_{\tau}-\mu \right) b(\tau)\\
& +\frac{U}{2}\int_0^{\beta} \d \tau b^{\ast}(\tau)b^{\ast}(\tau)b(\tau)b(\tau)\\
& -\kappa\int_0^{\beta} \d \tau \mathbf{\Psi}^{\ast}\mathbf{b}(\tau)\\
& +\frac{1}{2}\int_0^{\beta} \d \tau \int_0^{\beta} \d \tau' \mathbf{b}^{\ast}(\tau)\mathbb{\Delta}(\tau-\tau')\mathbf{b}(\tau').
\end{aligned}
\end{equation}
Here $\beta=1/T$ is the inverse of the temperature $T$, $\kappa=\sum_{i} t_{ij}=zt$ is a geometrical parameter depending on a lattice type, and $\tau$ is the imaginary (Matsubara) time. We use the Nambu vector notation for operators $\hat{b}$ and complex variables $b$, i.e.,
\begin{equation}
\mathbf{\hat{b}}=\left(
\begin{array}{l}
\hat{b}\\
\hat{b}^{\dag}
\end{array}
 \right), \ \ \
 \mathbf{b}=\left(
\begin{array}{l}
b\\
b^{\ast}
\end{array}
 \right).
\end{equation}
Then the connected Green functions are defined by
\begin{equation}\label{local_green}
\begin{aligned}
\mathbb{G}(\tau)&=-\langle T_{\tau} \mathbf{\hat{b}}(\tau) \mathbf{\hat{b}}^{\dag}(0) \rangle^c\\
&=-\langle T_{\tau} \mathbf{\hat{b}}(\tau) \mathbf{\hat{b}}^{\dag}(0) \rangle+\langle \mathbf{\hat{b}}(\tau)\rangle \langle\mathbf{\hat{b}}^{\dag}(0) \rangle,
\end{aligned}
\end{equation}
where $\langle\ldots\rangle$ denotes the equilibrium average in the grand canonical ensemble, and $\mathbb{G}(\beta+\tau)=\mathbb{G}(\tau)$.\cite{Negele} The last two terms in Eq. \eqref{loc_action} represent the coupling of the site to two types of external mean-fields: (i) the Bose-Einstein condensate (BEC) which is represented by a static mean-field $\mathbf{\Psi}$, and (ii) the dynamical mean-field of normal bosons represented by the matrix $\mathbb{\Delta}(\tau-\tau')$, the elements of which are hybridization functions.

The hybridization functions are related to the Green functions $\mathbb{G}^{(0)}$ for a lattice with a cavity at site $0$ (i.e., where site $0$ is removed) by\cite{byczuk_bdmft}
\begin{equation}
\mathbb{\Delta}(\tau-\tau')=\sum_{i,j\neq 0} t_{i0}t_{j0} \mathbb{G}^{(0)}_{ij}(\tau-\tau').
\end{equation}
Similarly, the condensate field $\mathbf{\Psi}$ for a lattice with a cavity at site $0$ reads
\begin{equation}
\mathbf{\Psi}=\langle \mathbf{\hat{b}}\rangle^{(0)}.
\end{equation}
The connection between the condensate field on a full lattice, $\mathbf{\Phi}= \langle \hat{\mathbf{b}}\rangle$, and the condensate field on a lattice with a cavity, $\mathbf{\Psi}$, is expressed by
\begin{equation}\label{self_con_phi}
\mathbf{\Psi}=\left[\mathbf{1}
+\frac{1}{\kappa}\int_0^{\beta} \d \tau \mathbb{\Delta}(\tau)\right]\mathbf{\Phi}.
\end{equation}

 Once the mean fields $\mathbf{\Psi}$ and $\mathbb{\Delta}(\tau)$ are known it is possible to solve the impurity problem and calculate the local Green functions and local condensate fields. This can be achieved by exact diagonalization\cite{hofstetter_2species_2009, dmft_ed}, the LCE approximation\cite{kauch_lce}, or CT-QMC.\cite{anders_dmft_short,anders_dmft} In this paper we use the CT-QMC method to solve the impurity problem. Our code is based on the work of Anders \emph{et al.}\cite{anders_dmft} We employ an improved calculation of the self-energy via two-particle Green functions, to be discussed in Section \ref{imp_SE} and in the Appendix. Tests of our code show numerical agreement with the results of Ref. \onlinecite{anders_dmft}.


\subsection{Self-consistency}

Once the single-particle Green functions have been obtained by solving the single-site problem, the local self-energy can be calculated from the Dyson equation
\begin{equation}\label{dyson_loc}
\mathbb{\Sigma}(\i\omega_n)=\begin{pmatrix}
\i\omega_n + \mu & 0\\
0 & -\i\omega_n+\mu
\end{pmatrix}
-\mathbb{\Delta}(\i\omega_n) - \left[ \mathbb{G}(\i\omega_n) \right]^{-1},
\end{equation}
where $\omega_n$ are even Matsubara frequencies for bosons ($\omega_n=2\pi n/\beta$). In the B-DMFT the self-energies are local, i.e., momentum independent. The momentum resolved Green functions are then given by
\begin{equation}\label{dyson_glob2}\begin{aligned}
 & \mathbb{G}(\mathbf{k},\i\omega_n)=\\
 & \left[\begin{pmatrix}
\i\omega_n+\mu-\epsilon_\mathbf{k} & 0\\
0 & -\i\omega_n+\mu-\epsilon_\mathbf{k}
\end{pmatrix}-\mathbb{\Sigma}(\i\omega_n)\right]^{-1},
\end{aligned}\end{equation}
where $\epsilon_\mathbf{k}$ is the dispersion relation for noninteracting bosons. To close the self-consistency equations we calculate the local Green functions from the $\mathbf{k}$-integrated Dyson equation
\begin{equation}\label{dyson_glob}\begin{aligned}
 & \mathbb{G}(\i\omega_n)= \sum_{\mathbf{k}} \mathbb{G}(\mathbf{k},\i\omega_n)=\\
& \int \d \epsilon D(\epsilon)\left[\begin{pmatrix}
\i\omega_n+\mu-\epsilon & 0\\
0 & -\i\omega_n+\mu-\epsilon
\end{pmatrix}-\mathbb{\Sigma}(\i\omega_n)\right]^{-1},
\end{aligned}\end{equation}
where $D(\epsilon)$ is the noninteracting density of states (DOS).


\subsection{Efficient calculation of self-energies}\label{imp_SE}

\begin{figure}[pt!]
\hspace{-0.7cm}
\resizebox{1.05\columnwidth}{!}{
\includegraphics{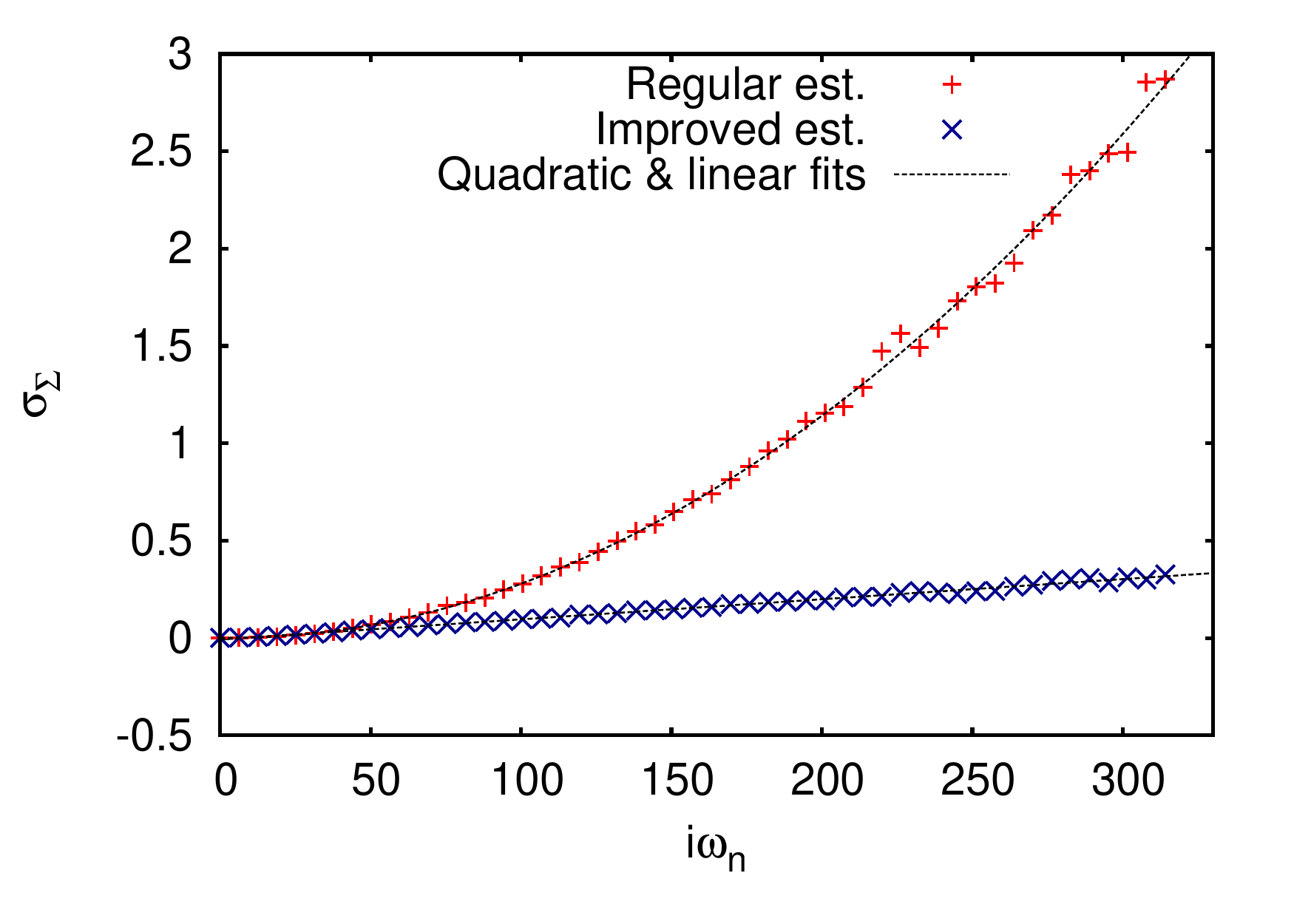}}
\caption{\label{self_energy} Comparison of the error of the real part of the self-energy calculated with the conventional (red) and the improved (blue) method. Parameters are $U=20 $, $\mu=0.23U$, $T =1 $. The improved method significantly reduces the error by making it linearly, rather than quadratically, dependent on frequency.}
\end{figure}%
The self-energies can be calculated by solving the single-impurity model \eqref{loc_action} with the CT-QMC method and then using Eq.~\eqref{dyson_loc}. However, this calculation is subject to stochastic uncertainties at high positive and negative frequencies for the following reason: In the first step the impurity problem is solved with the CT-QMC from which one obtains the Green function $G(\i\omega_n)$, which has the form $G(\i\omega_n)=\frac{1}{\i\omega_n} + O(\frac{1}{(\i\omega_n)^2})$ for large $|\omega_n|$. The numerical error of this result is almost independent of $\omega_n$. Therefore relative error grows linearly with $\omega_n$. In the second step the Green function is inverted and substituted into Eq.~\eqref{dyson_loc}. The linear term cancels out, and only terms of the order of $(\i\omega_n)^0$, $(\i\omega_n)^{-1}$ and smaller remain. This results in a progressively larger relative error, i.e., the error increases quadratically with frequency $\omega_n$ (see Fig.~\ref{self_energy}).

To improve the accuracy of the calculation of the self-energies we adapt the method proposed by Snoek and Hofstetter,\cite{dmft_snoek} and use the two-particle Green functions. The strategy goes as follows: Finding equations for the self-energies requires solving the equations of motion for the Green functions in the bosonic single-impurity model \eqref{loc_action}. This is similar to the method proposed for fermions.\cite{bulla} The final expression for the self-energy reads
\begin{equation}\label{imp_est}
\mathbb{\Sigma}(\i\omega_n)=\left[U\mathbb{F}(\i\omega_n) + (\kappa+\mu)\begin{pmatrix}
\phi\phi^{\ast} & \phi\phi \\
\phi^{\ast}\phi^{\ast} & \phi^{\ast}\phi
\end{pmatrix} \delta_{n0} \right] \mathbb{G}^{-1},
\end{equation}
where $\mathbb{F}(\i\omega_n)$ is the Fourier transform of a matrix of two-particle, \textit{disconnected} Green functions $\mathbb{F}(\tau)$ given by
\begin{equation}
\mathbb{F}(\tau)=-\bigg\langle T_\tau
\arraycolsep=1.4pt\def\arraystretch{1.5}
\begin{pmatrix}
[\hat{b}^{\dag} \hat{b} \hat{b}](\tau) \hat{b}^{\dag}(0) & [\hat{b}^{\dag} \hat{b} \hat{b}](\tau) \hat{b}(0) \\
[\hat{b}^{\dag} \hat{b}^{\dag} \hat{b}](\tau) \hat{b}^{\dag}(0) & [\hat{b}^{\dag} \hat{b}^{\dag} \hat{b}](\tau) \hat{b}(0)
\end{pmatrix}\bigg\rangle.
\end{equation}
Details of this derivation are presented in Appendix~\ref{AppA}. In Fig.~\ref{self_energy} we see that the improved method significantly reduces the error, i.e., it now grows only linearly in frequency.


\subsection{Analytic continuation}

The main goal of this paper is to calculate the momentum resolved spectral function $A(\mathbf{k},\omega)$ and the momentum integrated spectral function $A(\omega)$, respectively, of the Bose-Hubbard model~\eqref{bh_ham}. From the CT-QMC impurity solver we obtain the Green functions and the self-energies in imaginary time or in Matsubara frequencies. Since these functions are analytic in the upper complex plane we can analytically continue them to the real axis. This involves an inversion of the Hilbert transform:
\begin{equation}
 G(\i\omega_n) =\int_{-\infty}^{\infty}\d\omega \frac{A(\omega)}{\i\omega_n-\omega},
\end{equation}
or
\begin{equation}
G(\tau) =\int_{-\infty}^{\infty}\d\omega \frac{A(\omega)\e^{-\tau\omega}}{1-\e^{-\beta\omega}}.
\end{equation}
It is well-known that this is a numerically ill-posed problem, in particular for noisy data obtained with CT-QMC. To make the analytic continuation tractable we use the maximum entropy (MaxEnt) method.\cite{skilling_bryan,bryan,jarrell}
We found that the MaxEnt procedure sometimes fails to converge\footnote{We use the code which was implemented by Jarrell and Gubernatis following Skilling and Bryan,\cite{skilling_bryan} and Bryan\cite{bryan}.}, especially in calculations of sharply peaked momentum resolved spectral functions $A(\mathbf{k},\omega)$. In these cases we use the \textit{historic} rather than the \textit{Bryan} version of MaxEnt.\footnote{In MaxEnt we use Bayesian inference to specify the probability for finding a spectral function under a constraint of the data. We then search for a maximum of this probability. At the same time there exists a parameter which controls the weight with which the entropic prior enters the equations. In \textit{historic} MaxEnt we decrease the value of this parameter until the spectral function reproduces the data within the uncertainty of measurement. In \textit{Bryan} MaxEnt we use Bayesian inference to estimate the most probable value of this parameter.}


\section{Results}\label{sect3}
\begin{figure}
\resizebox{1.04\columnwidth}{!}{
\includegraphics{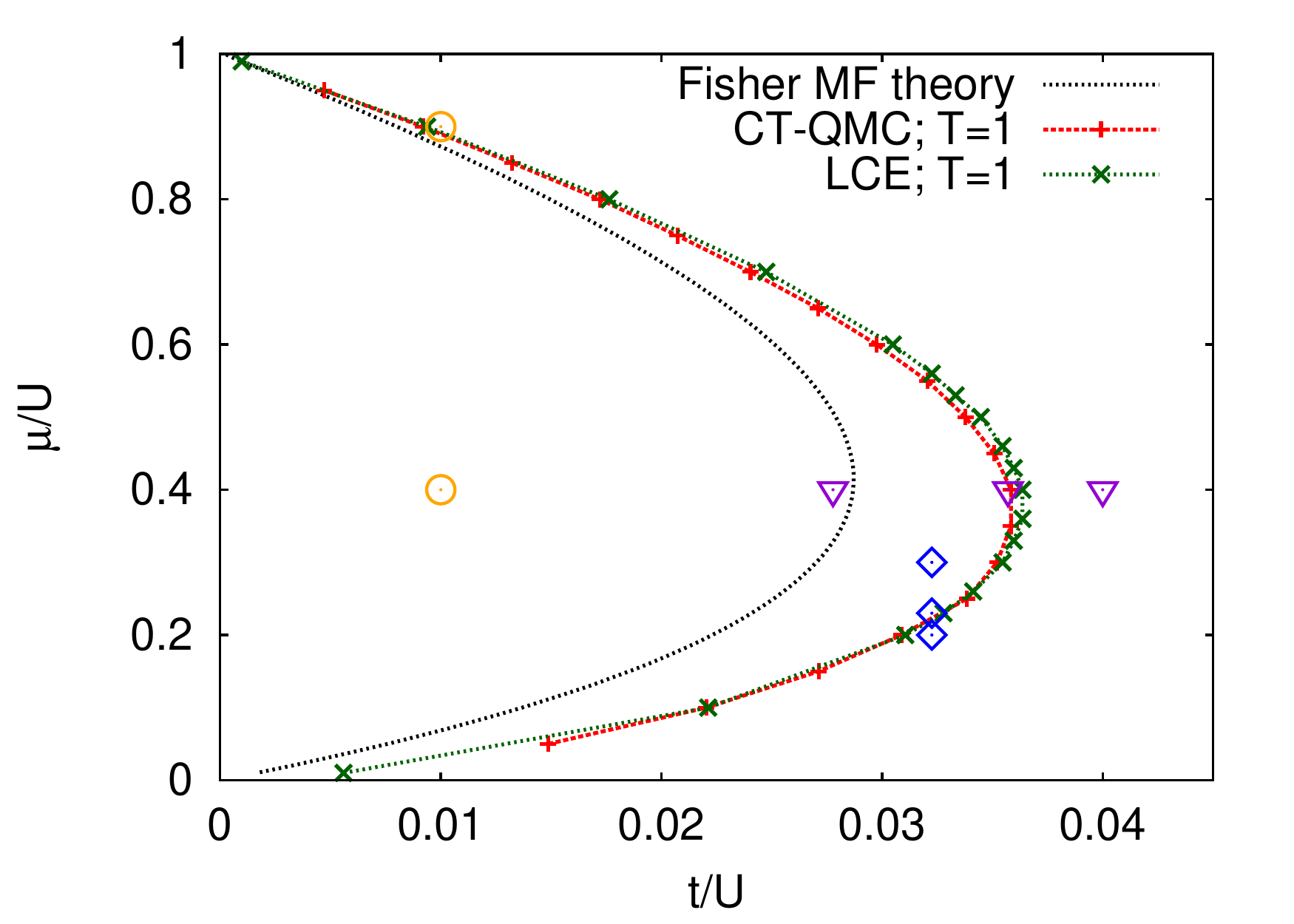}}
\caption{\label{phase_diag} Phase diagram of the Bose-Hubbard model on the simple cubic lattice obtained with the static Fisher mean-field (MF) theory\cite{fisher} and the B-DMFT, which is solved by CT-QMC and LCE. Only the first lobe corresponding to $\langle n\rangle \approx 1$ is plotted. Circles, diamonds and triangles represent sets of parameters at which calculations for strong and intermediate interactions, respectively, were performed. Circles (orange): Strong-coupling regime; triangles (violet) and diamonds (blue): Intermediate interaction regime. 
}
\end{figure}%
In the following we present results for spectral functions of the Bose-Hubbard model on a simple cubic lattice for weak, intermediate, and strong interaction strengths. For illustration we indicate the sets of parameters at which the calculations for strong and intermediate coupling, respectively, were performed in the phase diagram (Fig.~\ref{phase_diag}). The diagram was calculated within the static Fisher mean-field theory \cite{fisher} and the B-DMFT,\cite{byczuk_bdmft,anders_dmft_short} which was solved  by CT-QMC\cite{gull_ctqmc,anders_dmft} and LCE.\cite{kauch_lce}

%
\subsection{Weak-coupling limit}\label{non_lim}

\begin{figure}[pt!]
\hspace{-0.52cm}
\resizebox{1.05\columnwidth}{!}{
\includegraphics{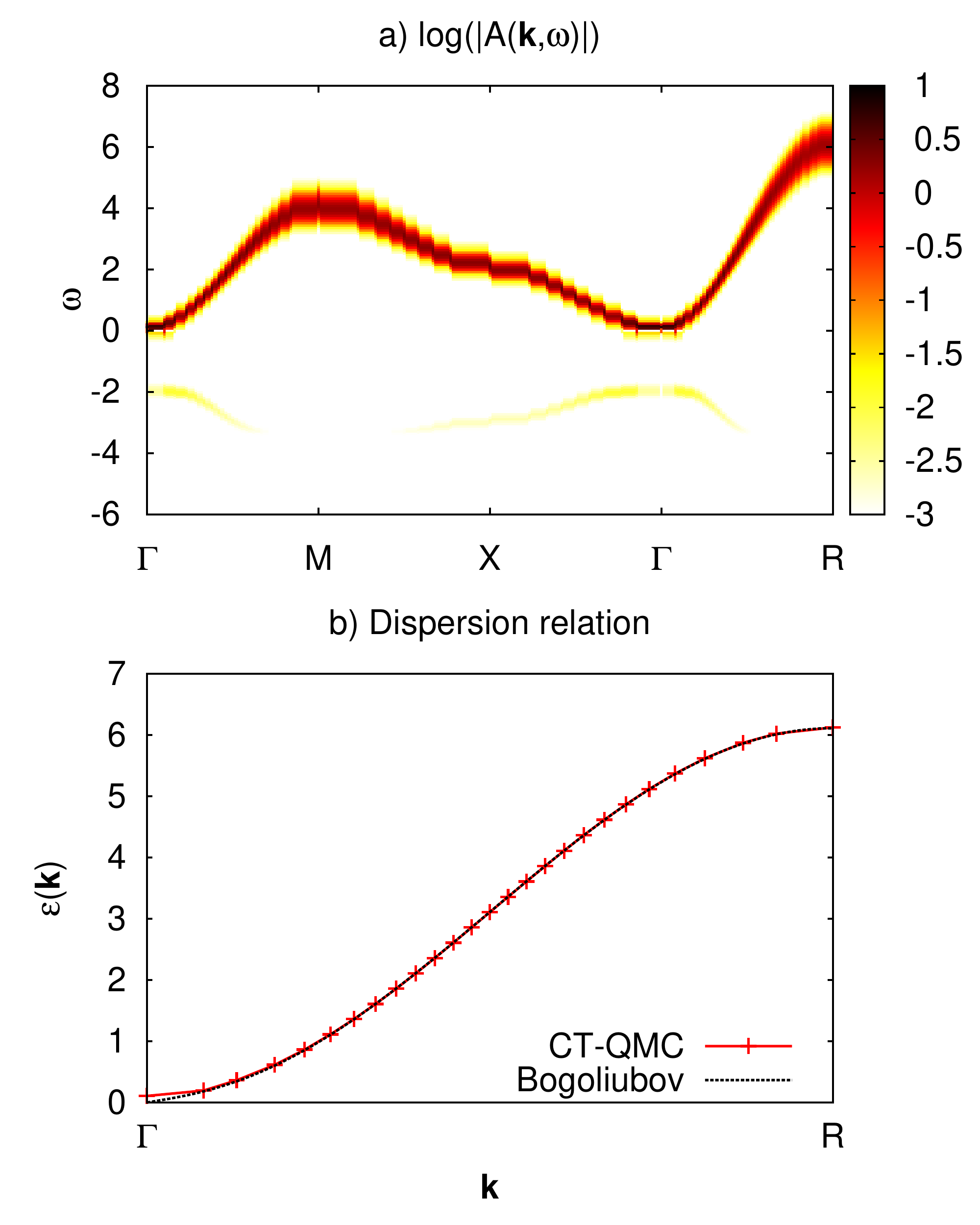}
}
\caption{\label{spectr_weak}(a) Momentum resolved spectral functions at $T=0.5$, $U=0.25$, and $\mu = -2.875$ along the symmetry lines in the first Brillouin zone of a simple cubic lattice. (b) Results obtained by B-DMFT (red) and by the Bogoliubov approximation (black) for the dispersion relations along the $\Gamma$-R line. }
\end{figure}%

Our results in the weak interaction limit were obtained for $T=0.5$, $U=0.25$, and $\mu = -2.875$. For these parameters the average occupation per site is $\langle n \rangle=0.4968\pm 0.0005$ and the average number of condensed bosons per site is $\langle b\rangle^2 = 0.4542 \pm 0.0004$. Since  $U\langle n\rangle \approx 0.1242 $ is small in comparison  to the bandwidth $2zt=6 $,  the weak-coupling Bogoliubov approximation\cite{bogoliubov,shi,wibg} can be expected to be applicable.  The momentum resolved spectral function $A(\mathbf{k},\omega)$  obtained within the B-DMFT is presented in Fig.~\ref{spectr_weak}a. There are two bands: with positive energies for particle addition and with negative energies for particle removal. Most of the spectral weight is concentrated in the upper band. The widths of the peaks for specific $\mathbf{k}$-points represent the mean lifetime of the quasiparticles. The weight of the lower band is orders of magnitude smaller, and therefore its 
exact shape and position is not determined reliably by MaxEnt. The dispersion relation $\epsilon(\mathbf{k})$, obtained from  $A(\mathbf{k},\omega)$ according to the definition
\begin{equation}\label{disprel_form}
\epsilon(\mathbf{k})=\max_\omega A(\mathbf{k},\omega),
\end{equation}
is shown in Fig.~\ref{spectr_weak}b. The dispersion relation obtained from the Bogoliubov approximation is found to be in very good agreement with the B-DMFT result, except near the $\Gamma$ point, where the B-DMFT dispersion does not go to zero. This is attributed to the fact that the Hugenholtz-Pines theorem\cite{hugenholtz} is not fulfilled in the B-DMFT as already reported by Anders \emph{et al.}\cite{anders_dmft}


\subsection{Strong-coupling limit}\label{sc_lim}

\begin{figure}[pt!]
\resizebox{1.02\columnwidth}{!}{
\includegraphics{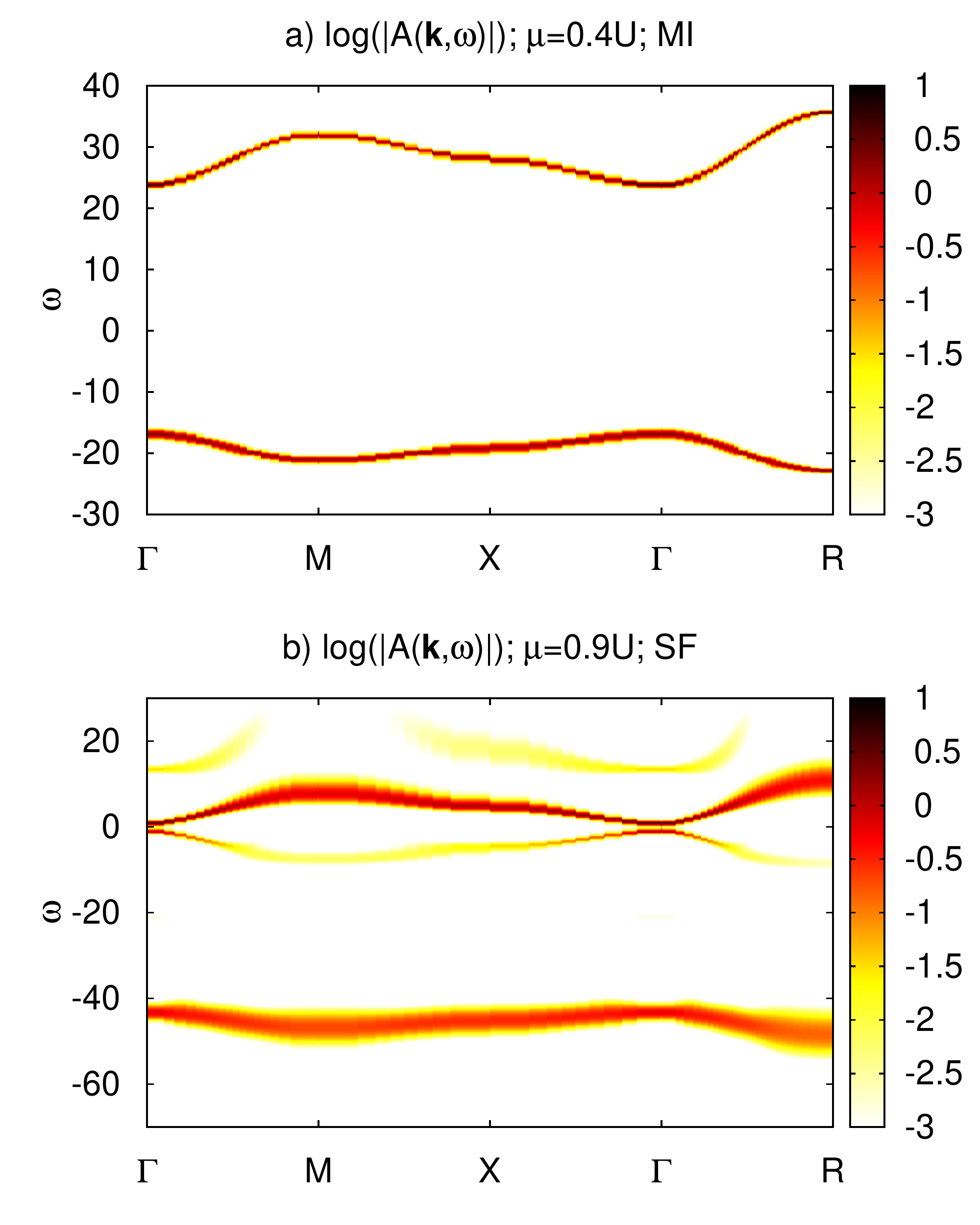}}
\caption{\label{spectr_SCL}Momentum resolved spectral function obtained from analytic continuation with MaxEnt of the CT-QMC data. Top panel: Mott insulating phase at $\mu=0.4U$ and $U=50$, bottom panel: Superfluid phase at $\mu=0.9U$ and $U=50$. Spectral functions are plotted along the symmetry lines in the first Brillouin zone for a simple cubic lattice.}
\end{figure}

Calculations in the limit of strong interactions were performed for $T=0.5$, and $U=50$. In Fig. \ref{spectr_SCL} we present the spectral function $A(\mathbf{k},\omega)$ for the Mott insulating phase at $\mu=0.4U$ and the superfluid phase at $\mu=0.9U$ (see Fig.~\ref{phase_diag}). A striking difference between the spectra in the Mott insulating and superfluid phases, respectively, is the number of bands and their width at specific $\mathbf{k}$-points.

In the Mott insulating phase there are two bands, separated by a gap whose width is approximately given by $U-3zt$. This value becomes exact in the atomic limit ($t\to 0$). The factor $3zt$ corresponds to the sum of half-widths of the upper and lower bands. The shape of this dispersion relation is almost the same as that for a noninteracting band. In the Mott insulating phase with average occupation $\langle n\rangle\approx 1$ one particle is frozen on every lattice site such that an extra particle or hole can move almost freely through the system. Since the energy for creating an excitation is large, the density of excitations is very low, and therefore one can neglect their interaction. As a result quasiparticle and quasihole excitations behave almost like noninteracting particles, i.e., have almost noninteracting dispersion relations. The width of the peaks is therefore small, implying that the mean lifetime is large. In fact, we checked that the width corresponds to the uncertainty of the analytic 
continuation  rather than to the mean lifetime.

The bandwidth of particle excitations is wider than that of hole excitations by approximately $2zt$. In the strong interaction limit the bandwidths of the hole and particle excitations in the Mott insulating phase with integer filling $\langle n \rangle$ are $2zt\langle n\rangle$ and $2zt(\langle n \rangle+1)$, respectively. This is a quantum effect related to particle indistinguishability, which is simple to derive by starting with a state in which each site is occupied by $n$ particles and treating the hopping term in the Hamiltonian~\eqref{bh_ham} as a small perturbation.
\begin{figure}[pt!]
\resizebox{1.05\columnwidth}{!}{
\includegraphics{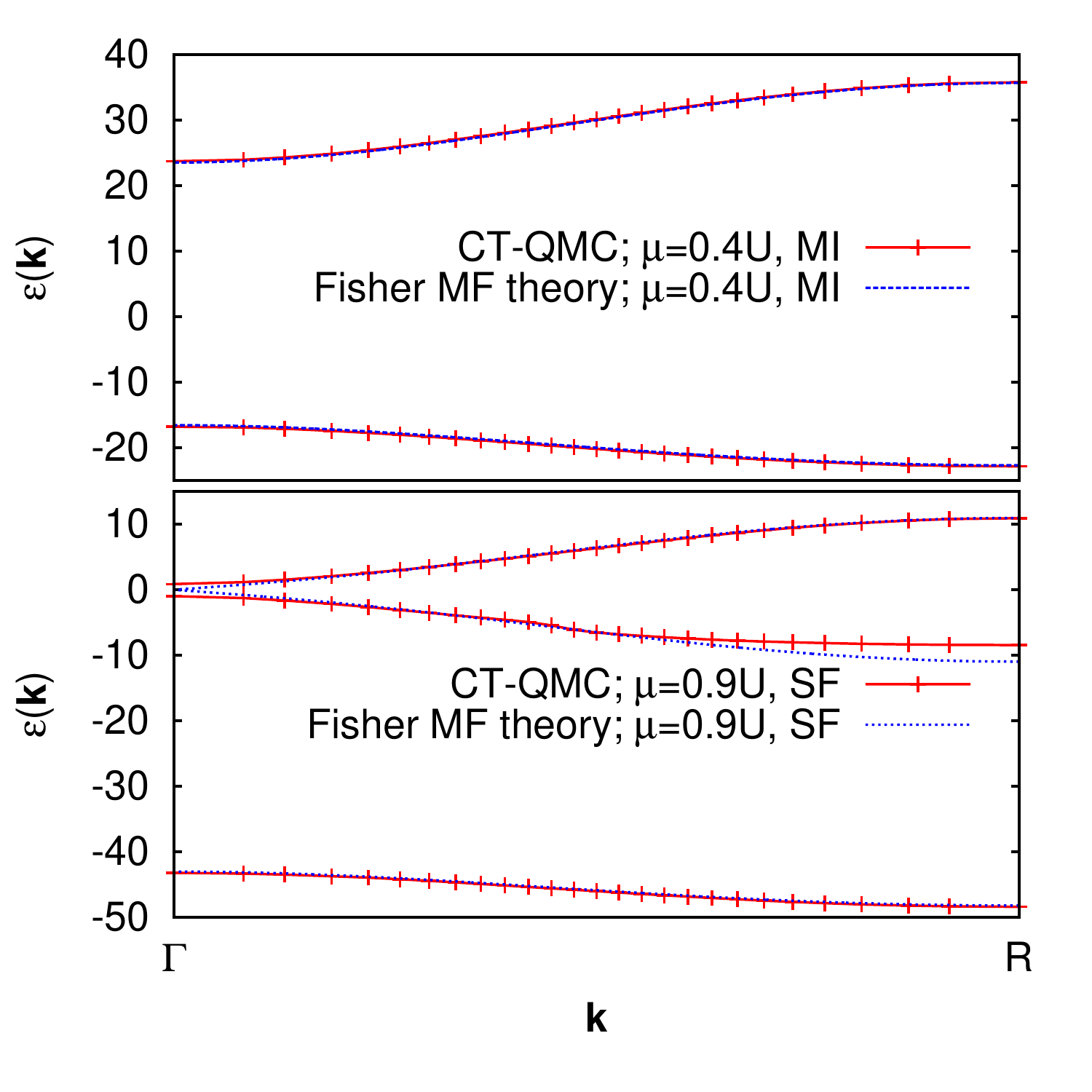}}
\caption{\label{disprel_SCL} Dispersion relation as calculated within the B-DMFT and the static Fisher mean-field (MF) theory\cite{fisher}, respectively. Top panel: Mott insulating phase at $\mu=0.4U$ and $U=50$; bottom panel: Superfluid phase at $\mu=0.9U$ and $U=50$. Only the three dominant bands are plotted.}
\end{figure}%

In the superfluid phase this is no longer valid. As the chemical potential increases, the energy for creating particle excitations decreases. Therefore the interaction between the quasiparticles needs to be taken into account. Indeed the spectrum of the superfluid is significantly different from that in the Mott insulating phase. In particular, we find four rather than two bands. Only the positions of the three lowest bands are determined reliably by MaxEnt. The four bands arise from two processes: (i) the splitting of a single band due to the interaction, similar to the Mott insulating phase; (ii) the creation of Bogoliubov quasiparticles due to the mixing of particle and hole excitations.  We checked that, in contrast to the Mott insulating phase, the width of the peaks for small $|\omega|$ is robust with respect to MaxEnt parameters as well as re-sampling and therefore represents the mean lifetime of quasiparticles and not the accuracy of analytic continuation.

In Fig.~\ref{disprel_SCL} we show the dispersion relations $\epsilon(\mathbf{k})$ obtained within the B-DMFT scheme according to Eq. \eqref{disprel_form}. We compare them with the dispersion relations obtained from the self-energies calculated within the static Fisher mean-field approximation.\cite{fisher} The results of both methods are in good agreement in the Mott insulating phase (Fig. \ref{disprel_SCL}, top panel).

In the superfluid phase (Fig.~\ref{disprel_SCL}, bottom panel) this comparison is presented for the three dominant bands. The results are also in good agreement with the static Fisher mean-field results.
In both approaches the high-energy (negative) band is similar to the band of a noninteracting hole.  The remaining two dispersions are linear for small values of $\mathbf{k}=0$ in the Fisher mean-field theory and correspond to massless Bogoliubov quasiparticles. In the B-DMFT we also see the linear behaviour except for the vicinity of the $\Gamma$-point. The dispersion relation around $\omega=0$ does not go to zero, since the Hugenholtz-Pines theorem is not obeyed in the B-DMFT (see also Section~\ref{non_lim}).  The slight deviation of the CT-QMC result from that of the static Fisher mean-field theory\cite{fisher} in the middle of the band around the R point is attributed to the finite resolution of MaxEnt.


\subsection{Intermediate interaction and comparison with the strong-coupling solver}\label{interm}
\begin{figure}[pt!]
\begin{center}
\resizebox{1.\columnwidth}{!}{
\includegraphics{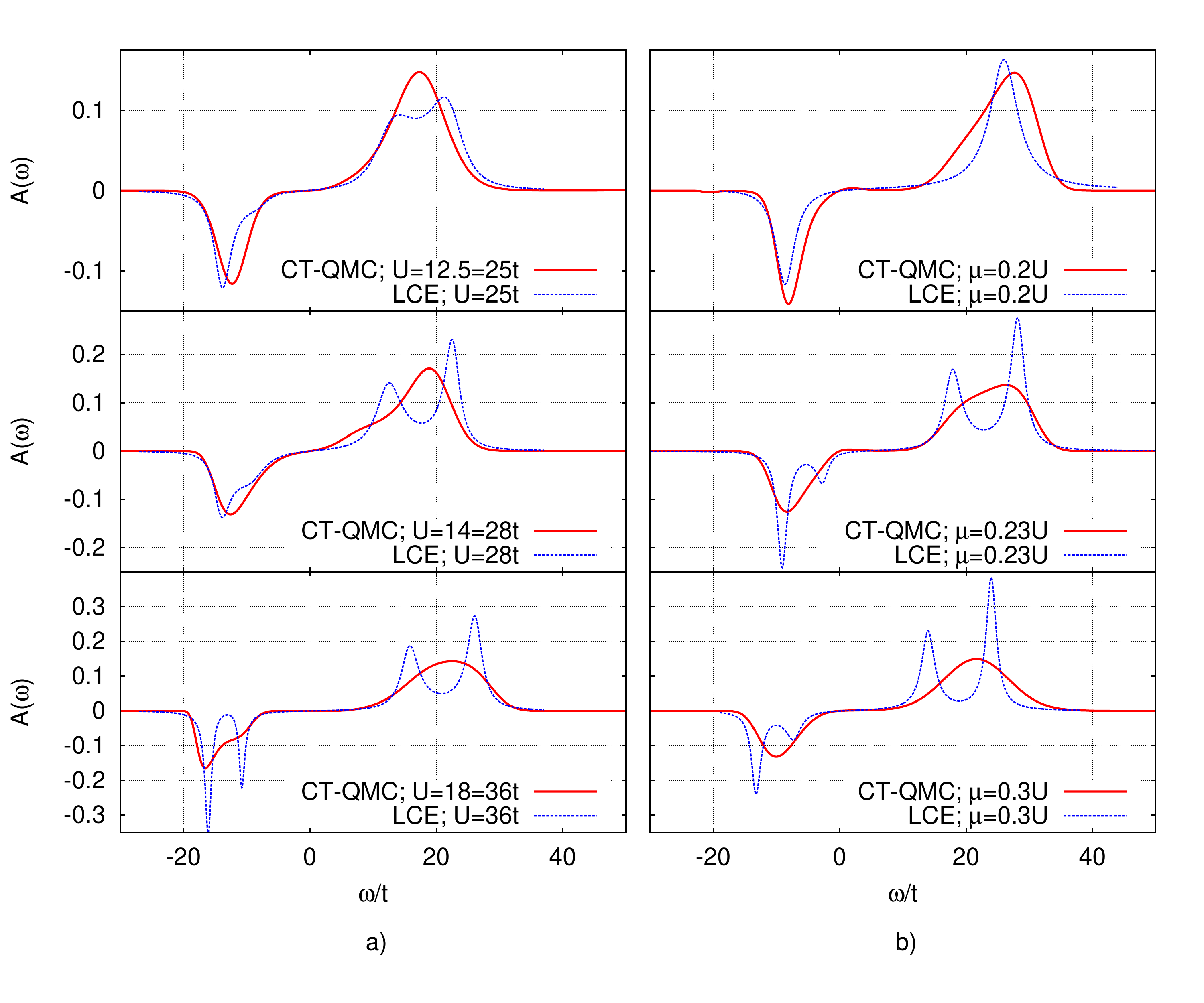}}
\caption{\label{lce_ctqmc} Spectral function $A(\omega)$ obtained with LCE (dotted line) and CT-QMC (full lines) at $T=1$. Pad\'e approximants were used for the data obtained with the LCE. Left panel: Calculations at $\mu=0.4U$ for different values of $U$; right panel: calculations at $U=15.5$ for different values of $\mu$. For $U=12.5$, $\mu=0.4U$ and $U=15.5$, $\mu=0.2U$ the system is in the superfluid phase; for reference see Fig.~\ref{phase_diag}.}
\end{center}
\end{figure}

\begin{figure}[hpt!]
\resizebox{1\columnwidth}{!}{
\includegraphics{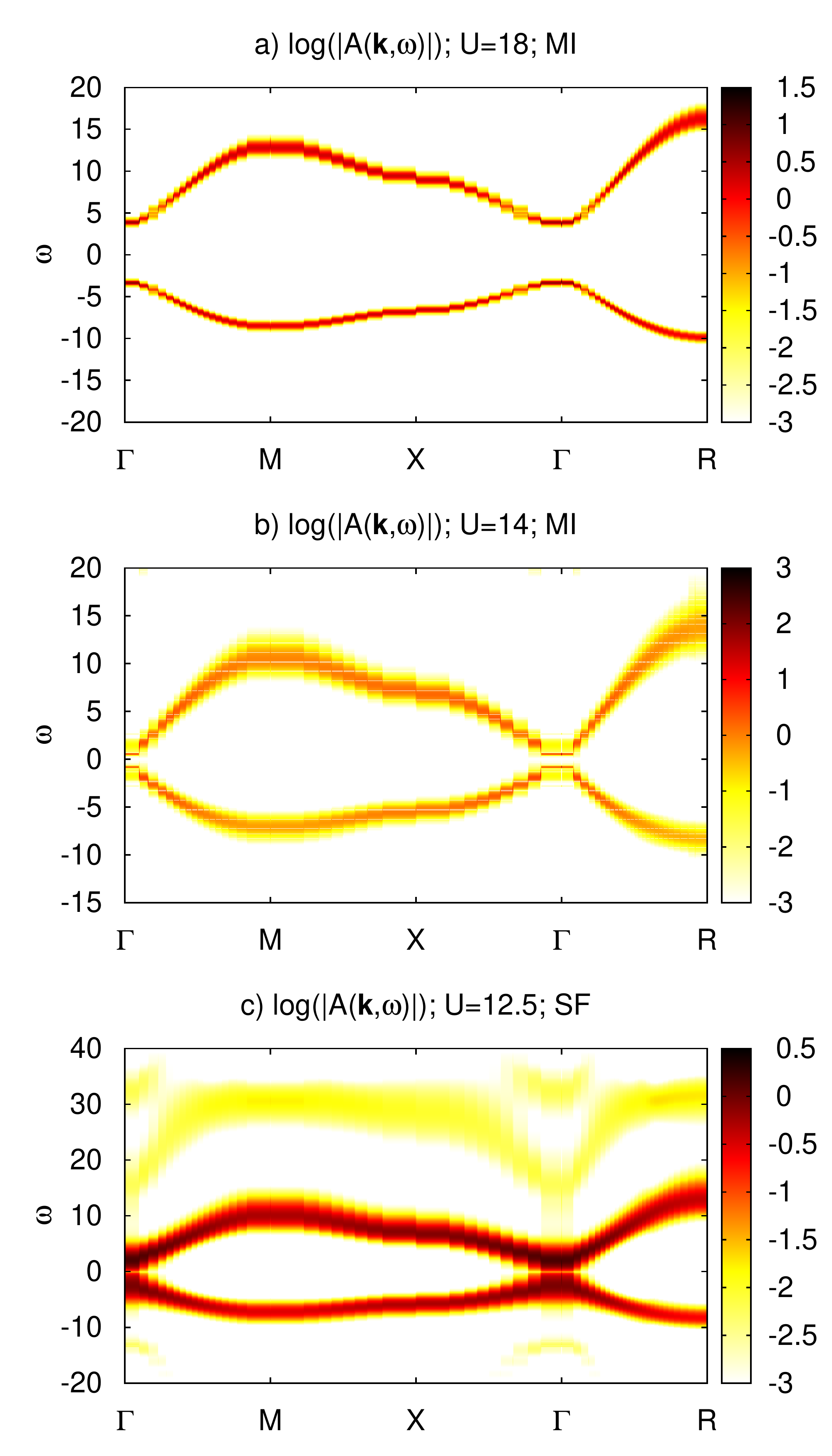}}
\caption{\label{Ak} Momentum resolved spectral function $A(\mathbf{k},\omega)$ in the first Brillouin zone of a simple cubic lattice at $\mu=0.4U$ and $T=1$. Interaction strengths are the same as in Fig.~\ref{lce_ctqmc}, left panel.}
\end{figure}%
\begin{figure}[h!]
\resizebox{1\columnwidth}{!}{
\includegraphics{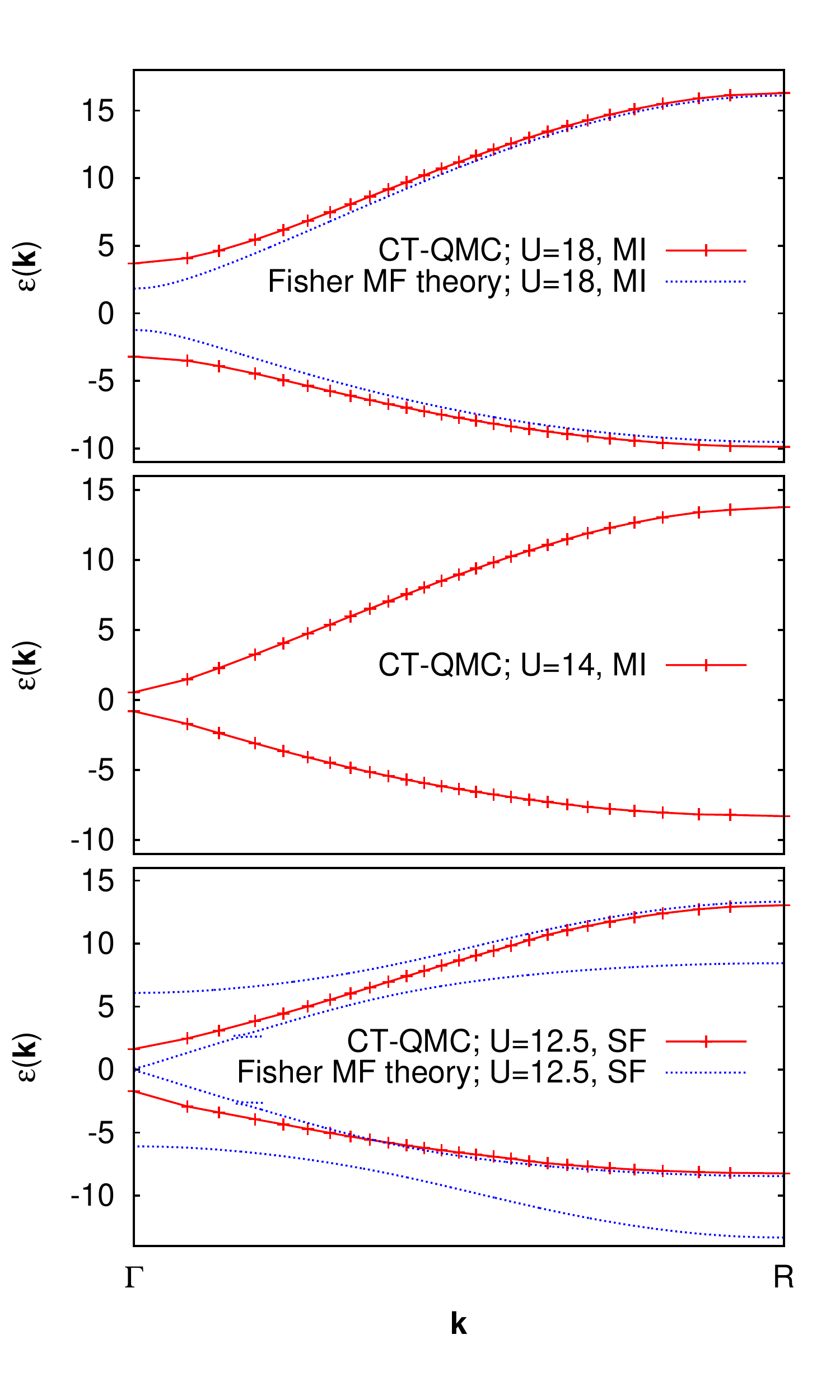}}
\caption{\label{disprel_int}Dispersion relation of correlated bosons for intermediate interaction strengths obtained within the B-DMFT. The momentum resolved spectral function $A(\mathbf{k},\omega)$ presented in Fig.~\ref{Ak} was employed together with Eq.~\eqref{disprel_form}. The results are compared to those of the static Fisher mean-field (MF) theory\cite{fisher}  except for the value $U=14$, where the B-DMFT finds the system to be in the Mott insulating phase, while it is in the superfluid phase according to the Fisher mean-field theory.}
\end{figure}%

The most interesting regime is that of intermediate interactions where both the Bogoliubov approximation and the static Fisher mean-field theory\cite{fisher} are no longer valid. We computed the spectral functions in this regime within B-DMFT and compare our CT-QMC results with those obtained within the LCE\cite{kauch_lce}.

We used the same set of parameters as in Ref. \onlinecite{kauch_lce}, i.e., each set corresponds to a point in a \{$\mu /U,t/U$\} parameter space for $T=1$. The selected points allow us to study the evolution of the spectral functions through the phase transition. Here we consider the phase transitions driven by the change of the interaction and by the change of the chemical potential (Fig~\ref{phase_diag}, left and right panel, respectively).

In Fig.\ref{lce_ctqmc} we present the B-DMFT results and compare with the LCE. There is good agreement regarding the widths and the positions of the bands in the superfluid phase. However, in the Mott insulating phase the LCE spectral functions develop a two-peak structure for both positive nad negative energies. This is not supported by the CT-QMC results. The origin of this feature in the LCE may be an overfitting of the numerical analytic continuation, since in the LCE the error is small but unknown.

We now focus on the momentum dependence of the spectral functions and the dispersion relations for the Mott insulating and superfluid phases (Figs.~\ref{Ak},~\ref{disprel_int}). The most striking feature is the difference of the width of the peaks for specific $\mathbf{k}$-points in these two phases. Deep in the Mott insulating phase the width represents the uncertainty of the MaxEnt fit. This is not the case in the superfluid phase in which the width represents the mean lifetime of quasiparticles. Effectively, deep in the insulating phase we can describe the particles as almost free, whereas approaching the phase transition the quasiparticles obtain a mean lifetime.

In the superfluid phase we observe high-energy excitations, which are shown in Fig.~\ref{Ak}c, resembling a structure found in the strong-coupling regime. Within the accuracy of our method a more precise calculation of their position is not possible.

The dispersion relations shown in Fig.\ref{disprel_int} are compared with the strong-coupling dispersions obtained for $U=18$ and $U=12.5$. In the Mott insulating phase the gap obtained by the static Fisher mean-field theory\cite{fisher} is much smaller than in the B-DMFT. Thereby the phase transition is shifted to larger interactions, see Fig~\ref{phase_diag}. In the superfluid phase we observe that, in contrast to the weak and strong-coupling limits, the dispersion relation obtained at intermediate coupling within the B-DMFT is not symmetric with respect to zero energy.\footnote{This is not visible in the weak-coupling results, since the negative energy band has a very low weight, such that its position is not determined reliably. However, the investigation of the self-energy\cite{anders_dmft} shows that it becomes static in the limit $U\to 0$, which leads to a symmetric dispersion relation.}
\begin{figure}[pt!]
\hspace{-1cm}
\resizebox{1.05\columnwidth}{!}{
\includegraphics{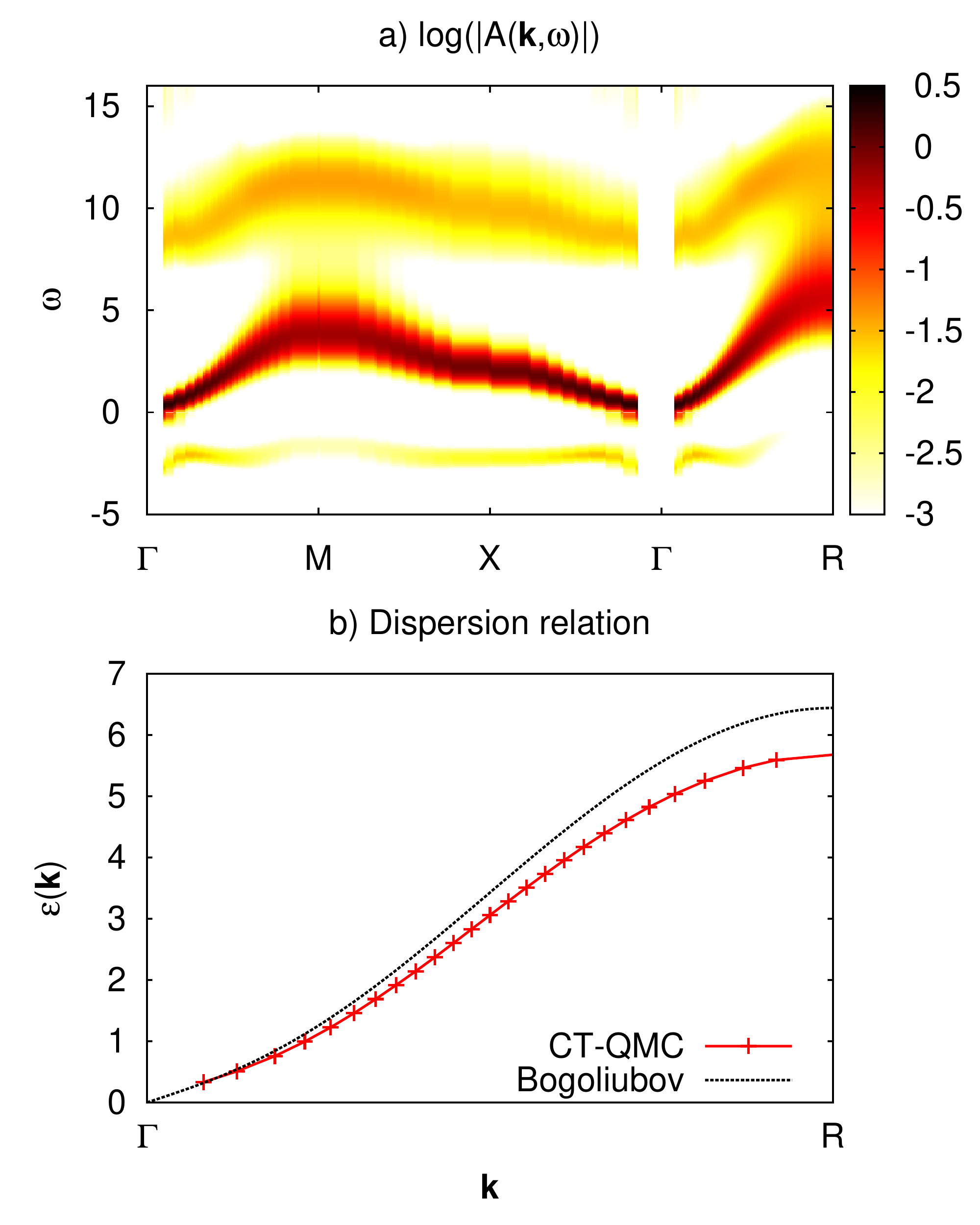}
}
\caption{\label{spectr_weak_U10}(a): Momentum resolved spectral function at $T=0.5$, $U=5$, and $\mu = -2.625$ along the symmetry lines in the first Brillouin zone of a simple cubic lattice. (b): Results obtained by the B-DMFT (red) and by the Bogoliubov approximation (black) for the dispersion relations along the $\Gamma$-R line.}
\end{figure}%

In Fig.~\ref{spectr_weak_U10} we present the results obtained in the dilute gas regime. The parameters are $T=0.5$, $U=5$, and $\mu=-2.625$. This corresponds to $\langle n \rangle = 0.12713\pm 0.00001$ and $\langle b \rangle = 0.30301\pm 0.00001$. As in the previous cases (superfluid phase) we observe that the peaks in the spectral function are wide, which means that the mean lifetime of excitations is finite. The negative energy  band has a weight which is orders of magnitude smaller, and therefore its exact shape and position is again not determined reliably by our method of analytic continuation. The results at the $\Gamma$-point are not included, since MaxEnt failed to converge for very small $\mathbf{k}$.

Although the particle density is low and the majority of the particles constitute the condensate, the interaction is not weak. Therefore one is not in the parameter range where the Bogoliubov approximation is applicable. Indeed, as seen in Fig.~\ref{spectr_weak_U10}b, the dispersion relation is not reproduced by this approximation. A similar discrepancy was reported in the experiment\cite{sengstock_nphys}, where at high momenta the Bogoliubov approximation was found to overestimate the measured excitation energy. A further effect of the interaction is the appearance of an additional band at high energies, Fig.~\ref{spectr_weak_U10}a. This high energy band appears to repel the one with low energies. As a result the low energy band is narrower. This explains the discrepancy between the B-DMFT and the Bogoliubov results, since the high energy band is absent in the latter.


\section{Summary}\label{sect4}

We presented a computational method which allows one to calculate spectral functions for bosonic systems described with the Bose-Hubbard Hamiltonian. This approach has the great advantage of being applicable for arbitrary values of $U/t$, the ratio of the interaction strength and the hopping amplitude. It reproduces well the results in the limiting cases of large and small values of the interaction. Results obtained by this method were shown for parameters where perturbative methods fail, thus providing new insights into the properties of correlated lattice bosons in previously inaccessible parameter ranges. The results show that while in the insulating phase particle and hole excitations tend to behave like free particles, in the superfluid phase the system is described by Bogoliubov quasiparticles with a finite mean lifetime.
Finally, we discussed an alternative method for calculating the local self-energy in the CT-QMC solver which employs two-particle Green functions, and which produces accurate data at large Matsubara frequencies.

\begin{acknowledgments}
This work was supported by the Foundation for Polish Science (FNP) International Ph.D. Projects Programme co-financed by the EU European Regional Development Fund. KB acknowledges support by the Foundation for Polish Science (FNP) through the TEAM/2010-6/2 project, co-financed by the EU European Regional Development Fund. This research was also supported in part by the Deutsche Forschungsgemeinschaft through TRR 80 (DV).
\end{acknowledgments}

\appendix

\section{Calculation of bosonic self-energy from equations of motion}\label{AppA}

We consider the Hamiltonian representation\cite{dmft_snoek} of the impurity model \eqref{loc_action}
\begin{equation}\label{app_ham}
\begin{array}{rl}
H_{imp}=&-\mu \hat{b}^{\dag} \hat{b} +\frac{U}{2} \hat{b}^{\dag}\hat{b}^{\dag} \hat{b} \hat{b} +\\
& \sum_l\left(\mathbf{\hat{b}}^\dag \mathbb{V}_{l} \mathbf{\hat{a}}_l + \epsilon_l \hat{a}_l^{\dag}  \hat{a}_l\right) - \kappa\mathbf{\Psi}^\dagger \mathbf{\hat{b}}
\end{array}
\end{equation}
where $\mathbf{\hat{a}}_l^\dag = (\hat{a}_l^\dag \ \hat{a}_l)$ and $\mathbf{\hat{b}}^\dag$ are vectors in Nambu notation. Here $\hat{a}_l$ ($\hat{a}^\dag_l$) annihilates (creates) a particle from the bath state $l$, $\hat{b}$ ($\hat{b}^\dag$) annihilates (creates) a particle on the impurity, $\mathbb{V}_{l}$ is the Nambu matrix of couplings between the impurity and a bath state $l$, $\epsilon_l$ is the energy of the bath state $l$, and the vector $\mathbf{\Psi}$ represents the condensate field to which the impurity is coupled.
The bath state energies $\epsilon_l$ and couplings $\mathbb{V}_{l}$ are chosen such that
\begin{equation}
\label{delta_def}
\mathbb{\Delta}(\i\omega_n) = \sum_l \mathbb{V}_l (\i\omega_n\bbsigma_3-\epsilon_l\mathbb{1})^{-1} \mathbb{V}_l^\dagger,
\end{equation}
where $\mathbb{\Delta}(\i\omega_n)$ is the dynamical mean field from Eq.~\eqref{loc_action}.

For this Hamiltonian we consider the disconnected Green functions
\begin{eqnarray}\label{app_G}
\mathbb{G}_{00}^d(\tau)=&-\langle T_\tau \mathbf{\hat{b}}(\tau) \mathbf{\hat{b}}^\dag (0) \rangle , \\
\mathbb{G}_{l0}^d(\tau)=&-\langle T_\tau \mathbf{\hat{a}}_l(\tau) \mathbf{\hat{b}}^\dag (0) \rangle .
\end{eqnarray}
Here the indices $0$ and $l$ refer to the impurity and bath states, respectively. The relation between $\mathbb{G}_{00}^d(\tau)$ and the connected impurity Green function $\mathbb{G}_{00}(\tau)$ defined in \eqref{local_green} is given by
\begin{equation}
\mathbb{G}^{d}_{00}(\tau)=\mathbb{G}_{00}(\tau)-\langle\mathbf{\hat{b}}(\tau)\rangle\langle\mathbf{\hat{b}}^{\dag}(0)\rangle.
\end{equation}

The time evolution of the operators is governed by the Hamiltonian $H_{imp}$ (for brevity we will drop the index henceforth), and we can calculate the derivatives of the Green functions with respect to imaginary time. For the local Green function we have
\begin{equation}\label{app_dtG00}
\arraycolsep=1.4pt\def\arraystretch{1.5}
\begin{array}{rcl}
\partial_\tau \mathbb{G}^d_{00}(\tau)& =& -\delta(\tau)\bbsigma_3 -\left\langle \begin{pmatrix}
[\hat{H},\hat{b}](\tau) \\ [\hat{H},\hat{b}^\dag ](\tau)
\end{pmatrix}
\begin{pmatrix}
\hat{b}^\dag (0) & \hat{b} (0)
\end{pmatrix}\right\rangle \\
&=&-\delta(\tau)\bbsigma_3+\mu \bbsigma_3 \mathbb{G}^d_{00}(\tau) -U\bbsigma_3 \mathbb{F}(\tau)\\
&&-\bbsigma_3\sum_l \mathbb{V}_{l}\mathbb{G}^d_{l0}(\tau)- \kappa\bbsigma_3\mathbf{\Psi}\mathbf{\Phi}^\dagger,
\end{array}
\end{equation}
where $[ \cdots,\cdots ]$ denotes the commutator, $\bbsigma_3$ is the diagonal Pauli matrix, $\mathbf{\Phi}^\dagger=(\langle b^\dag\rangle \ \langle b\rangle)$ is a Nambu vector, and the Nambu matrix $\mathbb{F}$ is defined as
\begin{equation}\label{app_F}
\begin{aligned}
& \mathbb{F}(\tau)= \\ 
& -\left\langle T_\tau
\arraycolsep=1.4pt\def\arraystretch{1.5}
\begin{pmatrix}
\hat{b}^\dag (\tau) \hat{b}(\tau) \hat{b}(\tau) \hat{b}^\dag (0) & \hat{b}^\dag (\tau) \hat{b}(\tau) \hat{b}(\tau) \hat{b}(0) \\
\hat{b}^\dag(\tau) \hat{b}^\dag(\tau) \hat{b} (\tau) \hat{b}^\dag (0) & \hat{b}^\dag(\tau) \hat{b}^\dag(\tau) \hat{b} (\tau) \hat{b} (0)
\end{pmatrix}\right\rangle;
\end{aligned}
\end{equation}
a similar relation holds for $\mathbb{G}^d_{l0}$:
\begin{equation}\label{app_dtGl0}
\arraycolsep=1.4pt\def\arraystretch{1.5}
\begin{array}{rl}
\partial_\tau \mathbb{G}^d_{l0}(\tau) &= -\left\langle \begin{pmatrix}
[\hat{H},\hat{a}_l](\tau) \\ [\hat{H},\hat{a}^\dag_l ](\tau)
\end{pmatrix}
\begin{pmatrix}
\hat{b}^\dag (0) & \hat{b} (0)
\end{pmatrix}\right\rangle \\
&=-\epsilon_l\bbsigma_3\mathbb{G}^d_{l0}-\bbsigma_3 \mathbb{V}^\dagger_{l}\mathbb{G}^d_{00}.
\end{array}
\end{equation}
To handle the imaginary time derivative we perform a Fourier transform to Matsubara frequencies and obtain
\begin{equation}\label{app_g_omn00}
\arraycolsep=1.4pt\def\arraystretch{1.5}
\begin{array}{rl}
\i\omega_n\bbsigma_3\mathbb{G}^d_{00}(\i\omega_n) =&  \mathbb{1}-\mu \mathbb{G}^d_{00}(\i\omega_n) + \sum_l \mathbb{V}_{l} \mathbb{G}^d_{l0} (\i\omega_n)\\
& +\delta_{n0}\kappa\mathbf{\Psi}\mathbf{\Phi}^\dagger +U\mathbb{F}(\i\omega_n),
\end{array}
\end{equation}
\begin{equation}\label{app_g_omn10}
\arraycolsep=1.4pt\def\arraystretch{1.5}
\begin{array}{rl}
\i\omega_n\bbsigma_3\mathbb{G}^d_{l0}(\i\omega_n) =
\epsilon_l\mathbb{G}^d_{l0}(\i\omega_n)+\mathbb{V}^\dagger_{l}\mathbb{G}^d_{00}(\i\omega_n).
\end{array}
\end{equation}
Using (\ref{app_g_omn10}) it is easy to find an expression for $\mathbb{G}^d_{l0}$ in terms of $\mathbb{G}^d_{00}$:
\begin{equation}\label{app_g_omn10_2}
\mathbb{G}^d_{l0}(\i\omega_n)=\left( \i\omega_n\bbsigma_3 - \epsilon_l\mathbb{1} \right)^{-1}\mathbb{V}^\dagger_{l}\mathbb{G}^d_{00}(\i\omega_n).
\end{equation}

We concentrate now on the case $n\neq 0$, i.e., nonzero Matsubara frequencies. Inserting \eqref{app_g_omn10_2} into \eqref{app_g_omn00} one obtains
\begin{equation}\label{app_int1}
\begin{aligned}
& \mathbb{G}^d_{00}(\i\omega_n)= \\
& \Big[ \i\omega_n\bbsigma_3  +\mu\mathbb{1}
- \sum_l \mathbb{V}_l (\i\omega_n\bbsigma_3-\epsilon_l\mathbb{1} )^{-1} \mathbb{V}_l^\dagger \Big]^{-1}(\mathbb{1}+U\mathbb{F}(\i\omega_n)).
\end{aligned}
\end{equation}
Using \eqref{delta_def} one can write
\begin{equation}\label{app_int2}
\mathbb{G}^d_{00}(\i\omega_n)= ( \i\omega_n\bbsigma_3  +\mu\mathbb{1}
-\mathbb{\Delta}(\i\omega_n))^{-1}(\mathbb{1}+U\mathbb{F}(\i\omega_n)).
\end{equation}
Combining \eqref{app_int2} with the Dyson equation \eqref{dyson_loc} we arrive at \eqref{imp_est} for non-zero Matsubara frequencies (for $n\neq0$ one has $\mathbb{G}^d_{00}=\mathbb{G}_{00}\equiv\mathbb{G}$).

By inserting \eqref{app_g_omn10_2} into \eqref{app_g_omn00} one finds in the case $n=0$:
\begin{equation}\label{app_int1_n0}
\mathbb{G}^d_{00}(0)= \Big(\mu\mathbb{1}
+ \sum_l  \epsilon_l^{-1} \mathbb{V}_l\mathbb{V}_l^\dagger \Big)^{-1}(\mathbb{1}+\kappa\mathbf{\Psi}\mathbf{\Phi}^\dagger+U\mathbb{F}(0)).
\end{equation}
Starting from \eqref{delta_def} for $n=0$ and replacing the disconnected Green function with the connected one leads to
\begin{equation}\label{app_int2_n0}
\mathbb{G}_{00}(0)-\mathbf{\Phi}\mathbf{\Phi}^\dagger= (\mu\mathbb{1}
- \mathbb{\Delta}(0) )^{-1}(\mathbb{1}+\kappa\mathbf{\Psi}\mathbf{\Phi}^\dagger+U\mathbb{F}(0)).
\end{equation}
In the last step the self-consistency condition \eqref{self_con_phi}, i.e., $\kappa\mathbf{\Psi}=(\kappa\mathbb{1}
+ \mathbb{\Delta}(0)) \mathbf{\Phi}$, is inserted into \eqref{app_int2_n0}. After some regrouping one obtains
\begin{equation}\begin{aligned}
& \mathbb{G}_{00}(0)-\mathbf{\Phi}\mathbf{\Phi}^\dagger= \\
&  (\mu\mathbb{1}
- \mathbb{\Delta}(0) )^{-1}[\mathbb{1}+(\kappa+\mu)\mathbf{\Phi}\mathbf{\Phi}^\dagger+U\mathbb{F}(0)]-\mathbf{\Phi}\mathbf{\Phi}^\dagger.
\end{aligned}
\end{equation}
After the above transformations one finally finds
\begin{equation}
\mathbb{G}_{00}(0)=(\mu\mathbb{1}
+ \mathbb{\Delta}(0) )^{-1}[\mathbb{1}+(\kappa+\mu)\mathbf{\Phi}\mathbf{\Phi}^\dagger+U\mathbb{F}(0)],
\end{equation}
which, in combination with the Dyson equation \eqref{dyson_loc}, reduces to \eqref{imp_est} for $n=0$.


\begin{thebibliography}{45}%
\makeatletter
\providecommand \@ifxundefined [1]{%
 \@ifx{#1\undefined}
}%
\providecommand \@ifnum [1]{%
 \ifnum #1\expandafter \@firstoftwo
 \else \expandafter \@secondoftwo
 \fi
}%
\providecommand \@ifx [1]{%
 \ifx #1\expandafter \@firstoftwo
 \else \expandafter \@secondoftwo
 \fi
}%
\providecommand \natexlab [1]{#1}%
\providecommand \enquote  [1]{``#1''}%
\providecommand \bibnamefont  [1]{#1}%
\providecommand \bibfnamefont [1]{#1}%
\providecommand \citenamefont [1]{#1}%
\providecommand \href@noop [0]{\@secondoftwo}%
\providecommand \href [0]{\begingroup \@sanitize@url \@href}%
\providecommand \@href[1]{\@@startlink{#1}\@@href}%
\providecommand \@@href[1]{\endgroup#1\@@endlink}%
\providecommand \@sanitize@url [0]{\catcode `\\12\catcode `\$12\catcode
  `\&12\catcode `\#12\catcode `\^12\catcode `\_12\catcode `\%12\relax}%
\providecommand \@@startlink[1]{}%
\providecommand \@@endlink[0]{}%
\providecommand \url  [0]{\begingroup\@sanitize@url \@url }%
\providecommand \@url [1]{\endgroup\@href {#1}{\urlprefix }}%
\providecommand \urlprefix  [0]{URL }%
\providecommand \Eprint [0]{\href }%
\providecommand \doibase [0]{http://dx.doi.org/}%
\providecommand \selectlanguage [0]{\@gobble}%
\providecommand \bibinfo  [0]{\@secondoftwo}%
\providecommand \bibfield  [0]{\@secondoftwo}%
\providecommand \translation [1]{[#1]}%
\providecommand \BibitemOpen [0]{}%
\providecommand \bibitemStop [0]{}%
\providecommand \bibitemNoStop [0]{.\EOS\space}%
\providecommand \EOS [0]{\spacefactor3000\relax}%
\providecommand \BibitemShut  [1]{\csname bibitem#1\endcsname}%
\let\auto@bib@innerbib\@empty
\bibitem [{\citenamefont {Greiner}\ \emph {et~al.}(2002)\citenamefont
  {Greiner}, \citenamefont {Mandel}, \citenamefont {Esslinger}, \citenamefont
  {H{\"a}nsch},\ and\ \citenamefont {Bloch}}]{bloch_short}%
  \BibitemOpen
  \bibfield  {author} {\bibinfo {author} {\bibfnamefont {M.}~\bibnamefont
  {Greiner}}, \bibinfo {author} {\bibfnamefont {O.}~\bibnamefont {Mandel}},
  \bibinfo {author} {\bibfnamefont {T.}~\bibnamefont {Esslinger}}, \bibinfo
  {author} {\bibfnamefont {T.~W.}\ \bibnamefont {H{\"a}nsch}}, \ and\ \bibinfo
  {author} {\bibfnamefont {I.}~\bibnamefont {Bloch}},\ }\href@noop {}
  {\bibfield  {journal} {\bibinfo  {journal} {Nature}\ }\textbf {\bibinfo
  {volume} {415}},\ \bibinfo {pages} {39} (\bibinfo {year} {2002})}\BibitemShut
  {NoStop}%
\bibitem [{\citenamefont {Bloch}\ \emph {et~al.}(2008)\citenamefont {Bloch},
  \citenamefont {Dalibard},\ and\ \citenamefont {Zwerger}}]{bloch_long}%
  \BibitemOpen
  \bibfield  {author} {\bibinfo {author} {\bibfnamefont {I.}~\bibnamefont
  {Bloch}}, \bibinfo {author} {\bibfnamefont {J.}~\bibnamefont {Dalibard}}, \
  and\ \bibinfo {author} {\bibfnamefont {W.}~\bibnamefont {Zwerger}},\
  }\href@noop {} {\bibfield  {journal} {\bibinfo  {journal} {Rev. Mod. Phys}\
  }\textbf {\bibinfo {volume} {80}},\ \bibinfo {pages} {885} (\bibinfo {year}
  {2008})}\BibitemShut {NoStop}%
\bibitem [{\citenamefont {Ozeri}\ \emph {et~al.}(2005)\citenamefont {Ozeri},
  \citenamefont {Katz}, \citenamefont {Steinhauer},\ and\ \citenamefont
  {Davidson}}]{bragg_rmp_2005}%
  \BibitemOpen
  \bibfield  {author} {\bibinfo {author} {\bibfnamefont {R.}~\bibnamefont
  {Ozeri}}, \bibinfo {author} {\bibfnamefont {N.}~\bibnamefont {Katz}},
  \bibinfo {author} {\bibfnamefont {J.}~\bibnamefont {Steinhauer}}, \ and\
  \bibinfo {author} {\bibfnamefont {N.}~\bibnamefont {Davidson}},\ }\href@noop
  {} {\bibfield  {journal} {\bibinfo  {journal} {Rev. Mod. Phys.}\ }\textbf
  {\bibinfo {volume} {77}},\ \bibinfo {pages} {187} (\bibinfo {year}
  {2005})}\BibitemShut {NoStop}%
\bibitem [{\citenamefont {Ernst}\ \emph {et~al.}(2010)\citenamefont {Ernst},
  \citenamefont {G\"{o}tze}, \citenamefont {Krauser}, \citenamefont {Pyka},
  \citenamefont {L\"uhmann}, \citenamefont {Pfannkuche},\ and\ \citenamefont
  {Sengstock}}]{sengstock_nphys}%
  \BibitemOpen
  \bibfield  {author} {\bibinfo {author} {\bibfnamefont {P.}~\bibnamefont
  {Ernst}}, \bibinfo {author} {\bibfnamefont {S.}~\bibnamefont {G\"{o}tze}},
  \bibinfo {author} {\bibfnamefont {J.~S.}\ \bibnamefont {Krauser}}, \bibinfo
  {author} {\bibfnamefont {K.}~\bibnamefont {Pyka}}, \bibinfo {author}
  {\bibfnamefont {D.}~\bibnamefont {L\"uhmann}}, \bibinfo {author}
  {\bibfnamefont {D.}~\bibnamefont {Pfannkuche}}, \ and\ \bibinfo {author}
  {\bibfnamefont {K.}~\bibnamefont {Sengstock}},\ }\href@noop {} {\bibfield
  {journal} {\bibinfo  {journal} {Nature Physics}\ }\textbf {\bibinfo {volume}
  {6}},\ \bibinfo {pages} {56} (\bibinfo {year} {2010})}\BibitemShut {NoStop}%
\bibitem [{\citenamefont {Papp}\ \emph {et~al.}(2008)\citenamefont {Papp},
  \citenamefont {Pino}, \citenamefont {Wild}, \citenamefont {Ronen},
  \citenamefont {Wieman}, \citenamefont {Jin},\ and\ \citenamefont
  {Cornell}}]{bragg_prl_2008}%
  \BibitemOpen
  \bibfield  {author} {\bibinfo {author} {\bibfnamefont {S.~B.}\ \bibnamefont
  {Papp}}, \bibinfo {author} {\bibfnamefont {J.~M.}\ \bibnamefont {Pino}},
  \bibinfo {author} {\bibfnamefont {R.~J.}\ \bibnamefont {Wild}}, \bibinfo
  {author} {\bibfnamefont {S.}~\bibnamefont {Ronen}}, \bibinfo {author}
  {\bibfnamefont {C.~E.}\ \bibnamefont {Wieman}}, \bibinfo {author}
  {\bibfnamefont {D.~S.}\ \bibnamefont {Jin}}, \ and\ \bibinfo {author}
  {\bibfnamefont {E.~A.}\ \bibnamefont {Cornell}},\ }\href@noop {} {\bibfield
  {journal} {\bibinfo  {journal} {Phys. Rev. Lett.}\ }\textbf {\bibinfo
  {volume} {101}},\ \bibinfo {pages} {135301} (\bibinfo {year}
  {2008})}\BibitemShut {NoStop}%
\bibitem [{\citenamefont {Du}\ \emph {et~al.}(2010)\citenamefont {Du},
  \citenamefont {Wan}, \citenamefont {Yesilada}, \citenamefont {Ryu},
  \citenamefont {Heinzen}, \citenamefont {Liang},\ and\ \citenamefont
  {Wu}}]{bragg_3d_njp_2010}%
  \BibitemOpen
  \bibfield  {author} {\bibinfo {author} {\bibfnamefont {X.}~\bibnamefont
  {Du}}, \bibinfo {author} {\bibfnamefont {S.}~\bibnamefont {Wan}}, \bibinfo
  {author} {\bibfnamefont {E.}~\bibnamefont {Yesilada}}, \bibinfo {author}
  {\bibfnamefont {C.}~\bibnamefont {Ryu}}, \bibinfo {author} {\bibfnamefont
  {D.~J.}\ \bibnamefont {Heinzen}}, \bibinfo {author} {\bibfnamefont
  {Z.}~\bibnamefont {Liang}}, \ and\ \bibinfo {author} {\bibfnamefont
  {B.}~\bibnamefont {Wu}},\ }\href@noop {} {\bibfield  {journal} {\bibinfo
  {journal} {New J. Phys.}\ }\textbf {\bibinfo {volume} {12}},\ \bibinfo
  {pages} {083025} (\bibinfo {year} {2010})}\BibitemShut {NoStop}%
\bibitem [{\citenamefont {Clement}\ \emph {et~al.}(2009)\citenamefont
  {Clement}, \citenamefont {Fabbri}, \citenamefont {Fallani}, \citenamefont
  {Fort},\ and\ \citenamefont {Inguscio}}]{bragg_MI_prl_2009}%
  \BibitemOpen
  \bibfield  {author} {\bibinfo {author} {\bibfnamefont {D.}~\bibnamefont
  {Clement}}, \bibinfo {author} {\bibfnamefont {N.}~\bibnamefont {Fabbri}},
  \bibinfo {author} {\bibfnamefont {L.}~\bibnamefont {Fallani}}, \bibinfo
  {author} {\bibfnamefont {C.}~\bibnamefont {Fort}}, \ and\ \bibinfo {author}
  {\bibfnamefont {M.}~\bibnamefont {Inguscio}},\ }\href@noop {} {\bibfield
  {journal} {\bibinfo  {journal} {Phys. Rev. Lett.}\ }\textbf {\bibinfo
  {volume} {102}},\ \bibinfo {pages} {155301} (\bibinfo {year}
  {2009})}\BibitemShut {NoStop}%
\bibitem [{\citenamefont {Fabbri}\ \emph {et~al.}(2012)\citenamefont {Fabbri},
  \citenamefont {Huber}, \citenamefont {Clement}, \citenamefont {Fallani},
  \citenamefont {Fort}, \citenamefont {Inguscio},\ and\ \citenamefont
  {Altman}}]{bragg_fabbri_prl_2012}%
  \BibitemOpen
  \bibfield  {author} {\bibinfo {author} {\bibfnamefont {N.}~\bibnamefont
  {Fabbri}}, \bibinfo {author} {\bibfnamefont {S.~D.}\ \bibnamefont {Huber}},
  \bibinfo {author} {\bibfnamefont {D.}~\bibnamefont {Clement}}, \bibinfo
  {author} {\bibfnamefont {L.}~\bibnamefont {Fallani}}, \bibinfo {author}
  {\bibfnamefont {C.}~\bibnamefont {Fort}}, \bibinfo {author} {\bibfnamefont
  {M.}~\bibnamefont {Inguscio}}, \ and\ \bibinfo {author} {\bibfnamefont
  {E.}~\bibnamefont {Altman}},\ }\href@noop {} {\bibfield  {journal} {\bibinfo
  {journal} {Phys. Rev. Lett.}\ }\textbf {\bibinfo {volume} {109}},\ \bibinfo
  {pages} {055301} (\bibinfo {year} {2012})}\BibitemShut {NoStop}%
\bibitem [{\citenamefont {Sengupta}\ and\ \citenamefont
  {Dupuis}(2005)}]{sengupta}%
  \BibitemOpen
  \bibfield  {author} {\bibinfo {author} {\bibfnamefont {K.}~\bibnamefont
  {Sengupta}}\ and\ \bibinfo {author} {\bibfnamefont {N.}~\bibnamefont
  {Dupuis}},\ }\href@noop {} {\bibfield  {journal} {\bibinfo  {journal} {Phys.
  Rev. A}\ }\textbf {\bibinfo {volume} {71}},\ \bibinfo {pages} {033629}
  (\bibinfo {year} {2005})}\BibitemShut {NoStop}%
\bibitem [{\citenamefont {Dupuis}(2009{\natexlab{a}})}]{dupuis_prl_2009}%
  \BibitemOpen
  \bibfield  {author} {\bibinfo {author} {\bibfnamefont {N.}~\bibnamefont
  {Dupuis}},\ }\href@noop {} {\bibfield  {journal} {\bibinfo  {journal} {Phys.
  Rev. Lett.}\ }\textbf {\bibinfo {volume} {102}},\ \bibinfo {pages} {190401}
  (\bibinfo {year} {2009}{\natexlab{a}})}\BibitemShut {NoStop}%
\bibitem [{\citenamefont {Dupuis}(2009{\natexlab{b}})}]{dupuis_pra_2009}%
  \BibitemOpen
  \bibfield  {author} {\bibinfo {author} {\bibfnamefont {N.}~\bibnamefont
  {Dupuis}},\ }\href@noop {} {\bibfield  {journal} {\bibinfo  {journal} {Phys.
  Rev. A}\ }\textbf {\bibinfo {volume} {80}},\ \bibinfo {pages} {043627}
  (\bibinfo {year} {2009}{\natexlab{b}})}\BibitemShut {NoStop}%
\bibitem [{\citenamefont {Sinner}\ \emph {et~al.}(2009)\citenamefont {Sinner},
  \citenamefont {Hasselmann},\ and\ \citenamefont
  {Kopietz}}]{kopietz_prl_2009}%
  \BibitemOpen
  \bibfield  {author} {\bibinfo {author} {\bibfnamefont {A.}~\bibnamefont
  {Sinner}}, \bibinfo {author} {\bibfnamefont {N.}~\bibnamefont {Hasselmann}},
  \ and\ \bibinfo {author} {\bibfnamefont {P.}~\bibnamefont {Kopietz}},\
  }\href@noop {} {\bibfield  {journal} {\bibinfo  {journal} {Phys. Rev. Lett.}\
  }\textbf {\bibinfo {volume} {102}},\ \bibinfo {pages} {120601} (\bibinfo
  {year} {2009})}\BibitemShut {NoStop}%
\bibitem [{\citenamefont {Sinner}\ \emph {et~al.}(2010)\citenamefont {Sinner},
  \citenamefont {Hasselmann},\ and\ \citenamefont
  {Kopietz}}]{kopietz_pra_2010}%
  \BibitemOpen
  \bibfield  {author} {\bibinfo {author} {\bibfnamefont {A.}~\bibnamefont
  {Sinner}}, \bibinfo {author} {\bibfnamefont {N.}~\bibnamefont {Hasselmann}},
  \ and\ \bibinfo {author} {\bibfnamefont {P.}~\bibnamefont {Kopietz}},\
  }\href@noop {} {\bibfield  {journal} {\bibinfo  {journal} {Phys. Rev. A}\
  }\textbf {\bibinfo {volume} {82}},\ \bibinfo {pages} {063632} (\bibinfo
  {year} {2010})}\BibitemShut {NoStop}%
\bibitem [{\citenamefont {Knap}\ \emph {et~al.}(2011)\citenamefont {Knap},
  \citenamefont {Arrigoni},\ and\ \citenamefont {von~der
  Linden}}]{knap_sf_2011}%
  \BibitemOpen
  \bibfield  {author} {\bibinfo {author} {\bibfnamefont {M.}~\bibnamefont
  {Knap}}, \bibinfo {author} {\bibfnamefont {E.}~\bibnamefont {Arrigoni}}, \
  and\ \bibinfo {author} {\bibfnamefont {W.}~\bibnamefont {von~der Linden}},\
  }\href@noop {} {\bibfield  {journal} {\bibinfo  {journal} {Phys. Rev. B}\
  }\textbf {\bibinfo {volume} {83}},\ \bibinfo {pages} {134507} (\bibinfo
  {year} {2011})}\BibitemShut {NoStop}%
\bibitem [{\citenamefont {Knap}\ \emph
  {et~al.}(2010{\natexlab{a}})\citenamefont {Knap}, \citenamefont {Arrigoni},\
  and\ \citenamefont {von~der Linden}}]{knap_1d_2010}%
  \BibitemOpen
  \bibfield  {author} {\bibinfo {author} {\bibfnamefont {M.}~\bibnamefont
  {Knap}}, \bibinfo {author} {\bibfnamefont {E.}~\bibnamefont {Arrigoni}}, \
  and\ \bibinfo {author} {\bibfnamefont {W.}~\bibnamefont {von~der Linden}},\
  }\href@noop {} {\bibfield  {journal} {\bibinfo  {journal} {Phys. Rev. B}\
  }\textbf {\bibinfo {volume} {81}},\ \bibinfo {pages} {235122} (\bibinfo
  {year} {2010}{\natexlab{a}})}\BibitemShut {NoStop}%
\bibitem [{\citenamefont {Knap}\ \emph
  {et~al.}(2010{\natexlab{b}})\citenamefont {Knap}, \citenamefont {Arrigoni},\
  and\ \citenamefont {von~der Linden}}]{knap_2d_2010}%
  \BibitemOpen
  \bibfield  {author} {\bibinfo {author} {\bibfnamefont {M.}~\bibnamefont
  {Knap}}, \bibinfo {author} {\bibfnamefont {E.}~\bibnamefont {Arrigoni}}, \
  and\ \bibinfo {author} {\bibfnamefont {W.}~\bibnamefont {von~der Linden}},\
  }\href@noop {} {\bibfield  {journal} {\bibinfo  {journal} {Phys. Rev. B}\
  }\textbf {\bibinfo {volume} {81}},\ \bibinfo {pages} {024301} (\bibinfo
  {year} {2010}{\natexlab{b}})}\BibitemShut {NoStop}%
\bibitem [{\citenamefont {Zaleski}(2012)}]{zaleski}%
  \BibitemOpen
  \bibfield  {author} {\bibinfo {author} {\bibfnamefont {T.~A.}\ \bibnamefont
  {Zaleski}},\ }\href@noop {} {\bibfield  {journal} {\bibinfo  {journal} {Phys.
  Rev. A}\ }\textbf {\bibinfo {volume} {85}},\ \bibinfo {pages} {043611}
  (\bibinfo {year} {2012})}\BibitemShut {NoStop}%
\bibitem [{\citenamefont {Zaleski}\ and\ \citenamefont
  {Kope\'c}(2014)}]{zaleski3d}%
  \BibitemOpen
  \bibfield  {author} {\bibinfo {author} {\bibfnamefont {T.~A.}\ \bibnamefont
  {Zaleski}}\ and\ \bibinfo {author} {\bibfnamefont {T.~K.}\ \bibnamefont
  {Kope\'c}},\ }\href@noop {} {\bibfield  {journal} {\bibinfo  {journal}
  {Physica B}\ }\textbf {\bibinfo {volume} {433}},\ \bibinfo {pages} {37}
  (\bibinfo {year} {2014})}\BibitemShut {NoStop}%
\bibitem [{\citenamefont {Kauch}\ \emph {et~al.}(2012)\citenamefont {Kauch},
  \citenamefont {Byczuk},\ and\ \citenamefont {Vollhardt}}]{kauch_lce}%
  \BibitemOpen
  \bibfield  {author} {\bibinfo {author} {\bibfnamefont {A.}~\bibnamefont
  {Kauch}}, \bibinfo {author} {\bibfnamefont {K.}~\bibnamefont {Byczuk}}, \
  and\ \bibinfo {author} {\bibfnamefont {D.}~\bibnamefont {Vollhardt}},\
  }\href@noop {} {\bibfield  {journal} {\bibinfo  {journal} {Phys. Rev. B}\
  }\textbf {\bibinfo {volume} {85}},\ \bibinfo {pages} {205115} (\bibinfo
  {year} {2012})}\BibitemShut {NoStop}%
\bibitem [{\citenamefont {Byczuk}\ and\ \citenamefont
  {Vollhardt}(2008)}]{byczuk_bdmft}%
  \BibitemOpen
  \bibfield  {author} {\bibinfo {author} {\bibfnamefont {K.}~\bibnamefont
  {Byczuk}}\ and\ \bibinfo {author} {\bibfnamefont {D.}~\bibnamefont
  {Vollhardt}},\ }\href@noop {} {\bibfield  {journal} {\bibinfo  {journal}
  {Phys. Rev. B}\ }\textbf {\bibinfo {volume} {77}},\ \bibinfo {pages} {235106}
  (\bibinfo {year} {2008})}\BibitemShut {NoStop}%
\bibitem [{\citenamefont {Metzner}\ and\ \citenamefont
  {Vollhardt}(1989)}]{metz_voll}%
  \BibitemOpen
  \bibfield  {author} {\bibinfo {author} {\bibfnamefont {W.}~\bibnamefont
  {Metzner}}\ and\ \bibinfo {author} {\bibfnamefont {D.}~\bibnamefont
  {Vollhardt}},\ }\href@noop {} {\bibfield  {journal} {\bibinfo  {journal}
  {Phys. Rev. Lett.}\ }\textbf {\bibinfo {volume} {62}},\ \bibinfo {pages}
  {324} (\bibinfo {year} {1989})}\BibitemShut {NoStop}%
\bibitem [{\citenamefont {Georges}\ \emph {et~al.}(1996)\citenamefont
  {Georges}, \citenamefont {Kotliar}, \citenamefont {Krauth},\ and\
  \citenamefont {Rozenberg}}]{kotliar_rmp}%
  \BibitemOpen
  \bibfield  {author} {\bibinfo {author} {\bibfnamefont {A.}~\bibnamefont
  {Georges}}, \bibinfo {author} {\bibfnamefont {G.}~\bibnamefont {Kotliar}},
  \bibinfo {author} {\bibfnamefont {W.}~\bibnamefont {Krauth}}, \ and\ \bibinfo
  {author} {\bibfnamefont {M.~J.}\ \bibnamefont {Rozenberg}},\ }\href@noop {}
  {\bibfield  {journal} {\bibinfo  {journal} {Rev. Mod. Phys.}\ }\textbf
  {\bibinfo {volume} {68}},\ \bibinfo {pages} {13} (\bibinfo {year}
  {1996})}\BibitemShut {NoStop}%
\bibitem [{\citenamefont {Anders}\ \emph {et~al.}(2011)\citenamefont {Anders},
  \citenamefont {Gull}, \citenamefont {Pollet}, \citenamefont {Troyer},\ and\
  \citenamefont {Werner}}]{anders_dmft}%
  \BibitemOpen
  \bibfield  {author} {\bibinfo {author} {\bibfnamefont {P.}~\bibnamefont
  {Anders}}, \bibinfo {author} {\bibfnamefont {E.}~\bibnamefont {Gull}},
  \bibinfo {author} {\bibfnamefont {L.}~\bibnamefont {Pollet}}, \bibinfo
  {author} {\bibfnamefont {M.}~\bibnamefont {Troyer}}, \ and\ \bibinfo {author}
  {\bibfnamefont {P.}~\bibnamefont {Werner}},\ }\href@noop {} {\bibfield
  {journal} {\bibinfo  {journal} {New J. Phys.}\ }\textbf {\bibinfo {volume}
  {13}},\ \bibinfo {pages} {075013} (\bibinfo {year} {2011})}\BibitemShut
  {NoStop}%
\bibitem [{\citenamefont {Hubener}\ \emph {et~al.}(2009)\citenamefont
  {Hubener}, \citenamefont {Snoek},\ and\ \citenamefont
  {Hofstetter}}]{hofstetter_2species_2009}%
  \BibitemOpen
  \bibfield  {author} {\bibinfo {author} {\bibfnamefont {A.}~\bibnamefont
  {Hubener}}, \bibinfo {author} {\bibfnamefont {M.}~\bibnamefont {Snoek}}, \
  and\ \bibinfo {author} {\bibfnamefont {W.}~\bibnamefont {Hofstetter}},\
  }\href@noop {} {\bibfield  {journal} {\bibinfo  {journal} {Phys. Rev. B}\
  }\textbf {\bibinfo {volume} {80}},\ \bibinfo {pages} {245109} (\bibinfo
  {year} {2009})}\BibitemShut {NoStop}%
\bibitem [{\citenamefont {Byczuk}\ and\ \citenamefont
  {Vollhardt}(2009)}]{byczuk_bf}%
  \BibitemOpen
  \bibfield  {author} {\bibinfo {author} {\bibfnamefont {K.}~\bibnamefont
  {Byczuk}}\ and\ \bibinfo {author} {\bibfnamefont {D.}~\bibnamefont
  {Vollhardt}},\ }\href@noop {} {\bibfield  {journal} {\bibinfo  {journal}
  {Ann. Phys.}\ }\textbf {\bibinfo {volume} {18}},\ \bibinfo {pages} {622}
  (\bibinfo {year} {2009})}\BibitemShut {NoStop}%
\bibitem [{\citenamefont {Anders}\ \emph {et~al.}(2012)\citenamefont {Anders},
  \citenamefont {Werner}, \citenamefont {Troyer}, \citenamefont {Sigrist},\
  and\ \citenamefont {Pollet}}]{anders_bf}%
  \BibitemOpen
  \bibfield  {author} {\bibinfo {author} {\bibfnamefont {P.}~\bibnamefont
  {Anders}}, \bibinfo {author} {\bibfnamefont {P.}~\bibnamefont {Werner}},
  \bibinfo {author} {\bibfnamefont {M.}~\bibnamefont {Troyer}}, \bibinfo
  {author} {\bibfnamefont {M.}~\bibnamefont {Sigrist}}, \ and\ \bibinfo
  {author} {\bibfnamefont {L.}~\bibnamefont {Pollet}},\ }\href@noop {}
  {\bibfield  {journal} {\bibinfo  {journal} {Phys. Rev. Lett.}\ }\textbf
  {\bibinfo {volume} {109}},\ \bibinfo {pages} {206401} (\bibinfo {year}
  {2012})}\BibitemShut {NoStop}%
\bibitem [{\citenamefont {Li}\ \emph {et~al.}(2011)\citenamefont {Li},
  \citenamefont {Bakhtiari}, \citenamefont {He},\ and\ \citenamefont
  {Hofstetter}}]{hofstetter_2species_trap_2011}%
  \BibitemOpen
  \bibfield  {author} {\bibinfo {author} {\bibfnamefont {Y.}~\bibnamefont
  {Li}}, \bibinfo {author} {\bibfnamefont {M.~R.}\ \bibnamefont {Bakhtiari}},
  \bibinfo {author} {\bibfnamefont {L.}~\bibnamefont {He}}, \ and\ \bibinfo
  {author} {\bibfnamefont {W.}~\bibnamefont {Hofstetter}},\ }\href@noop {}
  {\bibfield  {journal} {\bibinfo  {journal} {Phys. Rev. B}\ }\textbf {\bibinfo
  {volume} {84}},\ \bibinfo {pages} {144411} (\bibinfo {year}
  {2011})}\BibitemShut {NoStop}%
\bibitem [{\citenamefont {Strand}\ \emph {et~al.}(2015)\citenamefont {Strand},
  \citenamefont {Eckstein},\ and\ \citenamefont
  {Werner}}]{eckstein_noneq_2014}%
  \BibitemOpen
  \bibfield  {author} {\bibinfo {author} {\bibfnamefont {H.~U.~R.}\
  \bibnamefont {Strand}}, \bibinfo {author} {\bibfnamefont {M.}~\bibnamefont
  {Eckstein}}, \ and\ \bibinfo {author} {\bibfnamefont {P.}~\bibnamefont
  {Werner}},\ }\href@noop {} {\bibfield  {journal} {\bibinfo  {journal} {Phys.
  Rev. X}\ }\textbf {\bibinfo {volume} {5}},\ \bibinfo {pages} {011038}
  (\bibinfo {year} {2015})}\BibitemShut {NoStop}%
\bibitem [{\citenamefont {Gull}\ \emph {et~al.}(2011)\citenamefont {Gull},
  \citenamefont {Mills}, \citenamefont {Lichtenstein}, \citenamefont {Rubtsov},
  \citenamefont {Troyer},\ and\ \citenamefont {Werner}}]{gull_ctqmc}%
  \BibitemOpen
  \bibfield  {author} {\bibinfo {author} {\bibfnamefont {E.}~\bibnamefont
  {Gull}}, \bibinfo {author} {\bibfnamefont {A.~J.}\ \bibnamefont {Mills}},
  \bibinfo {author} {\bibfnamefont {A.~I.}\ \bibnamefont {Lichtenstein}},
  \bibinfo {author} {\bibfnamefont {A.~N.}\ \bibnamefont {Rubtsov}}, \bibinfo
  {author} {\bibfnamefont {M.}~\bibnamefont {Troyer}}, \ and\ \bibinfo {author}
  {\bibfnamefont {P.}~\bibnamefont {Werner}},\ }\href@noop {} {\bibfield
  {journal} {\bibinfo  {journal} {Rev. Mod. Phys.}\ }\textbf {\bibinfo {volume}
  {83}},\ \bibinfo {pages} {349} (\bibinfo {year} {2011})}\BibitemShut
  {NoStop}%
\bibitem [{\citenamefont {Anders}\ \emph {et~al.}(2010)\citenamefont {Anders},
  \citenamefont {Gull}, \citenamefont {Pollet}, \citenamefont {Troyer},\ and\
  \citenamefont {Werner}}]{anders_dmft_short}%
  \BibitemOpen
  \bibfield  {author} {\bibinfo {author} {\bibfnamefont {P.}~\bibnamefont
  {Anders}}, \bibinfo {author} {\bibfnamefont {E.}~\bibnamefont {Gull}},
  \bibinfo {author} {\bibfnamefont {L.}~\bibnamefont {Pollet}}, \bibinfo
  {author} {\bibfnamefont {M.}~\bibnamefont {Troyer}}, \ and\ \bibinfo {author}
  {\bibfnamefont {P.}~\bibnamefont {Werner}},\ }\href@noop {} {\bibfield
  {journal} {\bibinfo  {journal} {Phys. Rev. Lett.}\ }\textbf {\bibinfo
  {volume} {105}},\ \bibinfo {pages} {096402} (\bibinfo {year}
  {2010})}\BibitemShut {NoStop}%
\bibitem [{\citenamefont {Negele}\ and\ \citenamefont {Orland}(1988)}]{Negele}%
  \BibitemOpen
  \bibfield  {author} {\bibinfo {author} {\bibfnamefont {J.~W.}\ \bibnamefont
  {Negele}}\ and\ \bibinfo {author} {\bibfnamefont {H.}~\bibnamefont
  {Orland}},\ }\href@noop {} {\emph {\bibinfo {title} {Quantum Many-Particle
  Systems}}}\ (\bibinfo  {publisher} {Addison-Wesley},\ \bibinfo {address}
  {Melno Park},\ \bibinfo {year} {1988})\BibitemShut {NoStop}%
\bibitem [{\citenamefont {Hu}\ and\ \citenamefont {Tong}(2009)}]{dmft_ed}%
  \BibitemOpen
  \bibfield  {author} {\bibinfo {author} {\bibfnamefont {W.-J.}\ \bibnamefont
  {Hu}}\ and\ \bibinfo {author} {\bibfnamefont {N.~H.}\ \bibnamefont {Tong}},\
  }\href@noop {} {\bibfield  {journal} {\bibinfo  {journal} {Phys. Rev. B}\
  }\textbf {\bibinfo {volume} {80}},\ \bibinfo {pages} {245110} (\bibinfo
  {year} {2009})}\BibitemShut {NoStop}%
\bibitem [{\citenamefont {Snoek}\ and\ \citenamefont
  {Hofstetter}(2013)}]{dmft_snoek}%
  \BibitemOpen
  \bibfield  {author} {\bibinfo {author} {\bibfnamefont {M.}~\bibnamefont
  {Snoek}}\ and\ \bibinfo {author} {\bibfnamefont {W.}~\bibnamefont
  {Hofstetter}},\ }\enquote {\bibinfo {title} {Quantum gases: Finite
  temperature and non-equilibrium dynamics},}\ \ (\bibinfo  {publisher}
  {Imperial College Press, London},\ \bibinfo {year} {2013})\ Chap.~\bibinfo
  {chapter} {1}, pp.\ \bibinfo {pages} {355--365}\BibitemShut {NoStop}%
\bibitem [{\citenamefont {Bulla}\ \emph {et~al.}(1998)\citenamefont {Bulla},
  \citenamefont {Hewson},\ and\ \citenamefont {Pruschke}}]{bulla}%
  \BibitemOpen
  \bibfield  {author} {\bibinfo {author} {\bibfnamefont {R.}~\bibnamefont
  {Bulla}}, \bibinfo {author} {\bibfnamefont {A.~C.}\ \bibnamefont {Hewson}}, \
  and\ \bibinfo {author} {\bibfnamefont {T.}~\bibnamefont {Pruschke}},\
  }\href@noop {} {\bibfield  {journal} {\bibinfo  {journal} {J. Phys.: Condens.
  Mat.}\ }\textbf {\bibinfo {volume} {10}},\ \bibinfo {pages} {8365} (\bibinfo
  {year} {1998})}\BibitemShut {NoStop}%
\bibitem [{\citenamefont {Skilling}\ and\ \citenamefont
  {Bryan}(1984)}]{skilling_bryan}%
  \BibitemOpen
  \bibfield  {author} {\bibinfo {author} {\bibfnamefont {J.}~\bibnamefont
  {Skilling}}\ and\ \bibinfo {author} {\bibfnamefont {R.~K.}\ \bibnamefont
  {Bryan}},\ }\href@noop {} {\bibfield  {journal} {\bibinfo  {journal} {Mon.
  Not. R. astr. Soc.}\ }\textbf {\bibinfo {volume} {211}},\ \bibinfo {pages}
  {111} (\bibinfo {year} {1984})}\BibitemShut {NoStop}%
\bibitem [{\citenamefont {Bryan}(1990)}]{bryan}%
  \BibitemOpen
  \bibfield  {author} {\bibinfo {author} {\bibfnamefont {R.~K.}\ \bibnamefont
  {Bryan}},\ }\href@noop {} {\bibfield  {journal} {\bibinfo  {journal} {Eur.
  Biophys. J}\ }\textbf {\bibinfo {volume} {18}},\ \bibinfo {pages} {165}
  (\bibinfo {year} {1990})}\BibitemShut {NoStop}%
\bibitem [{\citenamefont {Jarrell}\ and\ \citenamefont
  {Gubernatis}(1996)}]{jarrell}%
  \BibitemOpen
  \bibfield  {author} {\bibinfo {author} {\bibfnamefont {M.}~\bibnamefont
  {Jarrell}}\ and\ \bibinfo {author} {\bibfnamefont {J.~E.}\ \bibnamefont
  {Gubernatis}},\ }\href@noop {} {\bibfield  {journal} {\bibinfo  {journal}
  {Phys. Rep.}\ }\textbf {\bibinfo {volume} {269}},\ \bibinfo {pages} {133}
  (\bibinfo {year} {1996})}\BibitemShut {NoStop}%
\bibitem [{Note1()}]{Note1}%
  \BibitemOpen
  \bibinfo {note} {We use the code which was implemented by Jarrell and
  Gubernatis following Skilling and Bryan,\cite {skilling_bryan} and Bryan\cite
  {bryan}.}\BibitemShut {Stop}%
\bibitem [{Note2()}]{Note2}%
  \BibitemOpen
  \bibinfo {note} {In MaxEnt we use Bayesian inference to specify the
  probability for finding a spectral function under a constraint of the data.
  We then search for a maximum of this probability. At the same time there
  exists a parameter which controls the weight with which the entropic prior
  enters the equations. In \protect \textit {historic} MaxEnt we decrease the
  value of this parameter until the spectral function reproduces the data
  within the uncertainty of measurement. In \protect \textit {Bryan} MaxEnt we
  use Bayesian inference to estimate the most probable value of this
  parameter.}\BibitemShut {Stop}%
\bibitem [{\citenamefont {Fisher}\ \emph {et~al.}(1989)\citenamefont {Fisher},
  \citenamefont {Weichman}, \citenamefont {Grinstein},\ and\ \citenamefont
  {Fisher}}]{fisher}%
  \BibitemOpen
  \bibfield  {author} {\bibinfo {author} {\bibfnamefont {M.~P.~A.}\
  \bibnamefont {Fisher}}, \bibinfo {author} {\bibfnamefont {P.~B.}\
  \bibnamefont {Weichman}}, \bibinfo {author} {\bibfnamefont {G.}~\bibnamefont
  {Grinstein}}, \ and\ \bibinfo {author} {\bibfnamefont {D.~S.}\ \bibnamefont
  {Fisher}},\ }\href@noop {} {\bibfield  {journal} {\bibinfo  {journal} {Phys.
  Rev. B}\ }\textbf {\bibinfo {volume} {40}},\ \bibinfo {pages} {546} (\bibinfo
  {year} {1989})}\BibitemShut {NoStop}%
\bibitem [{\citenamefont {Bogoliubov}(1947)}]{bogoliubov}%
  \BibitemOpen
  \bibfield  {author} {\bibinfo {author} {\bibfnamefont {N.~N.}\ \bibnamefont
  {Bogoliubov}},\ }\href@noop {} {\bibfield  {journal} {\bibinfo  {journal} {J.
  Phys. USSR}\ }\textbf {\bibinfo {volume} {23}},\ \bibinfo {pages} {11}
  (\bibinfo {year} {1947})}\BibitemShut {NoStop}%
\bibitem [{\citenamefont {Shi}\ and\ \citenamefont {Griffin}(1998)}]{shi}%
  \BibitemOpen
  \bibfield  {author} {\bibinfo {author} {\bibfnamefont {H.}~\bibnamefont
  {Shi}}\ and\ \bibinfo {author} {\bibfnamefont {A.}~\bibnamefont {Griffin}},\
  }\href@noop {} {\bibfield  {journal} {\bibinfo  {journal} {Phys. Rep.}\
  }\textbf {\bibinfo {volume} {304}},\ \bibinfo {pages} {1} (\bibinfo {year}
  {1998})}\BibitemShut {NoStop}%
\bibitem [{\citenamefont {Capogrosso-Sansone}\ \emph
  {et~al.}(2010)\citenamefont {Capogrosso-Sansone}, \citenamefont {Giorgini},
  \citenamefont {Pilati}, \citenamefont {Pollet}, \citenamefont {Prokof'ev},
  \citenamefont {Svistunov},\ and\ \citenamefont {Troyer}}]{wibg}%
  \BibitemOpen
  \bibfield  {author} {\bibinfo {author} {\bibfnamefont {B.}~\bibnamefont
  {Capogrosso-Sansone}}, \bibinfo {author} {\bibfnamefont {S.}~\bibnamefont
  {Giorgini}}, \bibinfo {author} {\bibfnamefont {S.}~\bibnamefont {Pilati}},
  \bibinfo {author} {\bibfnamefont {L.}~\bibnamefont {Pollet}}, \bibinfo
  {author} {\bibfnamefont {N.}~\bibnamefont {Prokof'ev}}, \bibinfo {author}
  {\bibfnamefont {B.}~\bibnamefont {Svistunov}}, \ and\ \bibinfo {author}
  {\bibfnamefont {M.}~\bibnamefont {Troyer}},\ }\href@noop {} {\bibfield
  {journal} {\bibinfo  {journal} {New J. Phys.}\ }\textbf {\bibinfo {volume}
  {12}},\ \bibinfo {pages} {043010} (\bibinfo {year} {2010})}\BibitemShut
  {NoStop}%
\bibitem [{\citenamefont {Hugenholtz}\ and\ \citenamefont
  {Pines}(1959)}]{hugenholtz}%
  \BibitemOpen
  \bibfield  {author} {\bibinfo {author} {\bibfnamefont {N.~M.}\ \bibnamefont
  {Hugenholtz}}\ and\ \bibinfo {author} {\bibfnamefont {D.}~\bibnamefont
  {Pines}},\ }\href@noop {} {\bibfield  {journal} {\bibinfo  {journal} {Phys.
  Rev.}\ }\textbf {\bibinfo {volume} {116}},\ \bibinfo {pages} {489} (\bibinfo
  {year} {1959})}\BibitemShut {NoStop}%
\bibitem [{Note3()}]{Note3}%
  \BibitemOpen
  \bibinfo {note} {This is not visible in the weak-coupling results, since the
  negative energy band has a very low weight, such that its position is not
  determined reliably. However, the investigation of the self-energy\cite
  {anders_dmft} shows that it becomes static in the limit $U\to 0$, which leads
  to a symmetric dispersion relation.}\BibitemShut {Stop}%
\end{thebibliography}


\end{document}